\newcommand{\DoPrePrint}{0} 

\ifnum\DoPrePrint=1
  \RequirePackage{lineno} 
  \documentclass[aps,prl,preprint,showpacs,superscriptaddress,nofootinbib,floatfix,linenumbers]{revtex4}

  \usepackage{lineno}
\else
  \documentclass[aps,prd,twocolumn,showpacs,superscriptaddress,nofootinbib]{revtex4}
\fi

\usepackage{amsmath}   
\usepackage{graphicx}  
\usepackage[caption=false,labelformat=simple]{subfig}
\usepackage{xspace}
\usepackage{units}
\usepackage{gensymb}
\usepackage{natbib}
\usepackage{hyperref}

\newcommand{\minerva}{MINERvA\xspace}
\newcommand{\minos}{MINOS\xspace}

\newcommand{\dune}{DUNE\xspace}
\newcommand{\genie}{\textsc{genie}\xspace}

\newcommand{\pknu}{\ensuremath{p \rightarrow K^{+} \nu}\xspace}
\newcommand{\dxs}{\ensuremath{d\sigma/dT_{K}}}

\newcommand{\numu}{\ensuremath{\nu_{\mu}}\xspace}

\newcommand{\numubar}{\ensuremath{\bar{\nu}_{\mu}}\xspace}

\newcommand{\pip}{\ensuremath{\pi^{+}}\xspace}

\newcommand{\piz}{\ensuremath{\pi^{0}}\xspace}
\newcommand{\lam}{\ensuremath{\Lambda}\xspace}

\newcommand{\kp}{\ensuremath{K^{+}}\xspace}

\newcommand{\km}{\ensuremath{K^{-}}\xspace}

\newcommand{\kz}{\ensuremath{K^{0}}\xspace}

\newcommand{\kzS}{\ensuremath{K^{0}_{\mathrm{S}}}\xspace}

\newcommand{\mup}{\ensuremath{\mu^{+}}\xspace}
\newcommand{\muminus}{\ensuremath{\mu^{-}}\xspace}
\newcommand{\mum}{\ensuremath{\mu^{-}}\xspace}

\newcommand{\sizecheck}{0} 
\newcommand{\PRLsupp}{0}   
\ifnum\PRLsupp=0
  
\else
\fi

\newif\ifpdf
\ifx\pdfoutput\undefined
   \pdffalse
\else
   \pdfoutput=1
   \pdftrue
\fi
\ifpdf
   \usepackage{graphicx}
   \usepackage{epstopdf}
\else
   \usepackage{graphicx}
\fi

\begin{document}

\ifnum\DoPrePrint=1
\linenumbers
\fi
\title{Measurement of $K^{+}$ production in charged-current $\nu_{\mu}$ interactions} 



\newcommand{\deceased}{Deceased}


\newcommand{\Rutgers}{Rutgers, The State University of New Jersey, Piscataway, New Jersey 08854, USA}
\newcommand{\Hampton}{Hampton University, Dept. of Physics, Hampton, VA 23668, USA}
\newcommand{\Dortmund}{Institute of Physics, Dortmund University, 44221, Germany }
\newcommand{\Otterbein}{Department of Physics, Otterbein University, 1 South Grove Street, Westerville, OH, 43081 USA}
\newcommand{\JMU}{James Madison University, Harrisonburg, Virginia 22807, USA}
\newcommand{\Florida}{University of Florida, Department of Physics, Gainesville, FL 32611}
\newcommand{\UCIrvine}{Department of Physics and Astronomy, University of California, Irvine, Irvine, California 92697-4575, USA}
\newcommand{\CBPF}{Centro Brasileiro de Pesquisas F\'{i}sicas, Rua Dr. Xavier Sigaud 150, Urca, Rio de Janeiro, Rio de Janeiro, 22290-180, Brazil}
\newcommand{\PUCP}{Secci\'{o}n F\'{i}sica, Departamento de Ciencias, Pontificia Universidad Cat\'{o}lica del Per\'{u}, Apartado 1761, Lima, Per\'{u}}
\newcommand{\INRM}{Institute for Nuclear Research of the Russian Academy of Sciences, 117312 Moscow, Russia}
\newcommand{\Jlab}{Jefferson Lab, 12000 Jefferson Avenue, Newport News, VA 23606, USA}
\newcommand{\Pittsburgh}{Department of Physics and Astronomy, University of Pittsburgh, Pittsburgh, Pennsylvania 15260, USA}
\newcommand{\Guanajuato}{Campus Le\'{o}n y Campus Guanajuato, Universidad de Guanajuato, Lascurain de Retana No. 5, Colonia Centro, Guanajuato 36000, Guanajuato M\'{e}xico.}
\newcommand{\Athens}{Department of Physics, University of Athens, GR-15771 Athens, Greece}
\newcommand{\Tufts}{Physics Department, Tufts University, Medford, Massachusetts 02155, USA}
\newcommand{\WM}{Department of Physics, College of William \& Mary, Williamsburg, Virginia 23187, USA}
\newcommand{\FNAL}{Fermi National Accelerator Laboratory, Batavia, Illinois 60510, USA}
\newcommand{\Purdue}{Department of Chemistry and Physics, Purdue University Calumet, Hammond, Indiana 46323, USA}
\newcommand{\MCLA}{Massachusetts College of Liberal Arts, 375 Church Street, North Adams, MA 01247}
\newcommand{\UMD}{Department of Physics, University of Minnesota -- Duluth, Duluth, Minnesota 55812, USA}
\newcommand{\Northwestern}{Northwestern University, Evanston, Illinois 60208}
\newcommand{\UNI}{Universidad Nacional de Ingenier\'{i}a, Apartado 31139, Lima, Per\'{u}}
\newcommand{\Rochester}{University of Rochester, Rochester, New York 14627 USA}
\newcommand{\Austin}{Department of Physics, University of Texas, 1 University Station, Austin, Texas 78712, USA}
\newcommand{\USM}{Departamento de F\'{i}sica, Universidad T\'{e}cnica Federico Santa Mar\'{i}a, Avenida Espa\~{n}a 1680 Casilla 110-V, Valpara\'{i}so, Chile}
\newcommand{\Geneva}{University of Geneva, 1211 Geneva 4, Switzerland}
\newcommand{\Chicago}{Enrico Fermi Institute, University of Chicago, Chicago, IL 60637 USA}
\newcommand{\hired}{}
\newcommand{\OregonState}{Department of Physics, Oregon State University, Corvallis, Oregon 97331, USA}
\newcommand{\oxford}{}
\newcommand{\bmeThanks}{now at SLAC National Accelerator Laboratory, Stanford, CA 94309, USA}
\newcommand{\higueraThanks}{now at University of Houston, Houston, TX 77204, USA}
\newcommand{\damartinezThanks}{now at Illinois Institute of Technology, Chicago, IL 60616, USA}
\newcommand{\joelmousseauThanks}{now at University of Michigan, Ann Arbor, MI 48109, USA}
\newcommand{\LazaThanks}{also at Department of Physics, University of Antananarivo, Madagascar}
\newcommand{\twaltonThanks}{now at Fermi National Accelerator Laboratory, Batavia, IL 60510, USA}
\newcommand{\jwolcottThanks}{Now at Tufts University, Medford, MA 02155, USA}
\newcommand{\mcgivernThanks}{now at Iowa State University, Ames, IA 50011, USA}

\author{C.M.~Marshall}                    \affiliation{\Rochester}
\author{L.~Aliaga}                        \affiliation{\WM}  \affiliation{\PUCP}
\author{O.~Altinok}                       \affiliation{\Tufts}
\author{L.~Bellantoni}                    \affiliation{\FNAL}
\author{A.~Bercellie}                     \affiliation{\Rochester}
\author{M.~Betancourt}                    \affiliation{\FNAL}
\author{A.~Bodek}                         \affiliation{\Rochester}
\author{A.~Bravar}                        \affiliation{\Geneva}
\author{H.~Budd}                          \affiliation{\Rochester}
\author{T.~Cai}                           \affiliation{\Rochester}
\author{M.F.~Carneiro}                    \affiliation{\CBPF}
\author{J.~Chvojka}                       \affiliation{\Rochester}
\author{H.~da~Motta}                      \affiliation{\CBPF}
\author{J.~Devan}                         \affiliation{\WM}
\author{S.A.~Dytman}                      \affiliation{\Pittsburgh}
\author{G.A.~D\'{i}az~}                   \affiliation{\Rochester}  \affiliation{\PUCP}
\author{B.~Eberly}\thanks{\bmeThanks}     \affiliation{\Pittsburgh}
\author{E.~Endress}                       \affiliation{\PUCP}
\author{J.~Felix}                         \affiliation{\Guanajuato}
\author{L.~Fields}                        \affiliation{\FNAL}  \affiliation{\Northwestern}
\author{A.~Filkins}                       \affiliation{\Rochester}
\author{R.~Fine}                          \affiliation{\Rochester}
\author{A.M.~Gago}                        \affiliation{\PUCP}
\author{R.~Galindo}                        \affiliation{\USM}
\author{H.~Gallagher}                     \affiliation{\Tufts}
\author{A.~Ghosh}                         \affiliation{\CBPF}
\author{T.~Golan}                         \affiliation{\Rochester}  \affiliation{\FNAL}
\author{R.~Gran}                          \affiliation{\UMD}
\author{S.~Griswold}                      \affiliation{\Rochester}
\author{D.A.~Harris}                      \affiliation{\FNAL}
\author{A.~Higuera}\thanks{\higueraThanks}  \affiliation{\Rochester}  \affiliation{\Guanajuato}
\author{K.~Hurtado}                       \affiliation{\CBPF}  \affiliation{\UNI}
\author{M.~Kiveni}                        \affiliation{\FNAL}
\author{J.~Kleykamp}                      \affiliation{\Rochester}
\author{M.~Kordosky}                      \affiliation{\WM}
\author{T.~Le}                            \affiliation{\Tufts}  \affiliation{\Rutgers}
\author{E.~Maher}                         \affiliation{\MCLA}
\author{I.~Majoros}                       \affiliation{\Otterbein}
\author{S.~Manly}                         \affiliation{\Rochester}
\author{W.A.~Mann}                        \affiliation{\Tufts}
\author{D.A.~Martinez~Caicedo}\thanks{\damartinezThanks}  \affiliation{\CBPF}
\author{K.S.~McFarland}                   \affiliation{\Rochester}  \affiliation{\FNAL}
\author{C.L.~McGivern}\thanks{\mcgivernThanks}       \affiliation{\Pittsburgh}
\author{A.M.~McGowan}                     \affiliation{\Rochester}
\author{B.~Messerly}                      \affiliation{\Pittsburgh}
\author{J.~Miller}                        \affiliation{\USM}
\author{A.~Mislivec}                      \affiliation{\Rochester}
\author{J.G.~Morf\'{i}n}                  \affiliation{\FNAL}
\author{J.~Mousseau}\thanks{\joelmousseauThanks}  \affiliation{\Florida}
\author{D.~Naples}                        \affiliation{\Pittsburgh}
\author{J.K.~Nelson}                      \affiliation{\WM}
\author{A.~Norrick}                       \affiliation{\WM}
\author{Nuruzzaman}                       \affiliation{\Rutgers}  \affiliation{\USM}
\author{J.~Osta}                          \affiliation{\FNAL}
\author{V.~Paolone}                       \affiliation{\Pittsburgh}
\author{J.~Park}                          \affiliation{\Rochester}
\author{C.E.~Patrick}                     \affiliation{\Northwestern}
\author{G.N.~Perdue}                      \affiliation{\FNAL}  \affiliation{\Rochester}
\author{L.~Rakotondravohitra}\thanks{\LazaThanks}  \affiliation{\FNAL}
\author{M.A.~Ramirez}                     \affiliation{\Guanajuato}
\author{R.D.~Ransome}                     \affiliation{\Rutgers}
\author{H.~Ray}                           \affiliation{\Florida}
\author{L.~Ren}                           \affiliation{\Pittsburgh}
\author{D.~Rimal}                         \affiliation{\Florida}
\author{P.A.~Rodrigues}                   \affiliation{\Rochester}
\author{M.~Rosenberg}                     \affiliation{\Pittsburgh}
\author{D.~Ruterbories}                   \affiliation{\Rochester}
\author{H.~Schellman}                     \affiliation{\OregonState}  \affiliation{\Northwestern}
\author{D.W.~Schmitz}                     \affiliation{\Chicago}  \affiliation{\FNAL}
\author{L.A.~Shadler}                       \affiliation{\Rochester}
\author{C.~Simon}                         \affiliation{\UCIrvine}
\author{C.J.~Solano~Salinas}              \affiliation{\UNI}
\author{S.F.~S\'{a}nchez~}                \affiliation{\PUCP}
\author{B.G.~Tice}                        \affiliation{\Rutgers}
\author{E.~Valencia}                      \affiliation{\Guanajuato}
\author{T.~Walton}\thanks{\twaltonThanks}  \affiliation{\Hampton}
\author{Z.~Wang}                          \affiliation{\Rochester}
\author{P.~Watkins}                       \affiliation{\Otterbein}
\author{K.~Wiley}                         \affiliation{\Rochester}
\author{J.~Wolcott}\thanks{\jwolcottThanks}  \affiliation{\Rochester}
\author{M.Wospakrik}                      \affiliation{\Florida}
\author{D.~Zhang}                         \affiliation{\WM}

\collaboration{ \minerva  Collaboration}\ \noaffiliation

\date{\today}

\begin{abstract}

Production of \kp mesons in charged-current \numu interactions on plastic scintillator (CH) is measured using \minerva exposed to the low-energy NuMI beam at Fermilab. Timing information is used to isolate a sample of 885 charged-current events containing a stopping \kp which decays at rest. The differential cross section in \kp kinetic energy, \dxs, is observed to be relatively flat between 0 and 500 MeV. Its shape is in good agreement with the prediction by the \genie neutrino event generator when final-state interactions are included, however the data rate is lower than the prediction by 15\%.

\end{abstract}
\ifnum\sizecheck=0  
\maketitle
\fi

\section{Introduction}
\label{sec:intro}

The energy spectrum of mesons produced in neutrino-nucleus interactions is modified by strong interactions with the residual nucleus. Recent high-statistics measurements of charged-current \pip production by MiniBooNE~\cite{miniboone_pion} and MINERvA~\cite{minerva_pion} have shown tension with available models~\cite{jan_pion}. A study of \kp production is complementary because of differences in the nuclear interaction due to strangeness conservation. Previous measurements of neutrino-induced charged-current \kp production have been carried out in bubble chambers with very limited statistics~\cite{anl2, anl12ft, bnl7ft, gargamelle}. We report the first high-statistics measurement of this process based on a sample of 1755 selected event candidates, of which 885 are estimated to be charged-current \kp events with $T_{K} < 600$~MeV.

At neutrino energies below 2 GeV, Cabibbo suppressed single kaon production $\nu_{\mu} N \rightarrow \mum \kp N$ is the dominant \kp production mechanism. At higher energies, \kp mesons arise via associated production accompanied by strangeness $= -1$~baryons ($\Lambda$, $\Sigma^{\pm}$) or mesons (\km, $\bar{K}^{0}$) such that there is no net change in strangeness ($\Delta S = 0$). This can occur through an intermediate resonance state or in deep inelastic scattering (DIS) by hadronization, the production of mesons and baryons from the struck quark. In particular, $s\bar{s}$ pairs created in hadronization lead to pairs of strange particles in the final state. 

Production of \kp by atmospheric neutrinos is a background in experimental searches for the proton decay \pknu, a channel favored by Grand Unification Theories which incorporate supersymmetry. The simplest minimal supersymmetric models~\cite{susygut1, susygut2} give proton lifetimes that have been excluded by experiment. However, other models~\cite{susygut3, susygut4, susygut5, susygut6, susygut7, susygut8} allow proton lifetimes greater than $10^{34}$ years, consistent with the current experimental lower bound of $5.6 \times 10^{33}$ years from a 260 kiloton-year exposure by Super-Kamiokande~\cite{skpknu}. The \kp from proton decay is below Cherenkov threshold in water, but a liquid argon time projection chamber such as DUNE~\cite{dune} is able to reconstruct the \kp momentum precisely. The \kp momentum spectrum in \pknu depends on the momentum distribution of the initial-state protons inside the nucleus. A related issue is the extent to which \kp mesons born inside the nucleus experience final-state interactions (FSI) as they emerge into the detector medium. Kaons produced by neutrinos are subject to the same interactions. Measuring \kp production by neutrinos on carbon is a first step toward understanding the spectrum for \pknu in the argon of the \dune far detector.

Kaon-nucleus and pion-nucleus reactions differ because of strangeness conservation. Absorption is the dominant feature in the pion-nucleus inelastic cross section at pion kinetic energies in the few 100s of MeV. In \km-nucleus scattering, the \km can be absorbed, converting a bound nucleon into a hyperon. The analogous process for \kp-nucleus scattering is forbidden because there are no antibaryons in the nucleus. A \kp produced inside the nucleus will exit unless it charge exchanges to a \kz. In addition, \kp can be produced in \pip-nucleus reactions by strong processes such as $\pi^{+} n \rightarrow K^{+} \Lambda$. In the Giessen Boltzmann-Uehling-Uhlenbeck model~\cite{gibuu}, this kind of reaction gives an enhancement to the \kp production cross section at low \kp momentum. In \genie~\cite{genieNIM}, the event generator used by \minerva and many other experiments, 13\% of \kp produced in carbon reinteract before exiting the nucleus, distorting the spectrum toward lower kaon energies. \genie does not include \kp production either by pions or charge exchange in its FSI model.

This paper reports a measurement at high statistics of inclusive charged-current \kp production by muon neutrinos, $\nu_{\mu}$ CH $\rightarrow \mum \kp X$. The differential cross section in \kp kinetic energy is measured and compared to predictions of current neutrino event generators with and without FSI treatments.

\section{\minerva Experiment}
\label{sec:expt}

\minerva is a dedicated neutrino-nucleus cross section experiment in the NuMI beamline~\cite{numi} at Fermilab. The detector consists of a core of strips of solid plastic scintillator ``tracker'' surrounded by calorimeters on the sides and downstream end. The electromagnetic and hadronic calorimeters intersperse scintillator with passive planes of lead and steel, respectively. The upstream nuclear targets region is used only to veto front-entering events for this result. The MINOS near detector is located 2~m downstream of MINERvA. Positive muons from antineutrino-induced charged-current reactions are rejected using curvature, but the muon momentum measurement is not used in this analysis.

The scintillator strips are arranged into planes stacked perpendicular to the horizontal axis, and are rotated $0^{\circ}$ and $\pm60^{\circ}$ with respect to the vertical axis to enable unambiguous three-dimensional tracking of charged particles. The cross section of the strips is triangular with a base edge of 3.4~cm and a height of 1.7~cm. In the center of each strip is a wavelength-shifting optical fiber which is mirrored at one end and read out by a 64-channel multi-anode photomultiplier tube at the other.

A hit is defined as an energy deposit in a single scintillator strip. The uncalibrated hit time is the time of the earliest charge recorded on a single channel, with an electronics resolution of 2.2~ns. When a charge threshold is exceeded, charge is integrated for 151~ns such that subsequent energy deposits in one strip due to the same neutrino interaction accumulate onto one hit. In particular, the timing of a delayed \kp decay product is lost if the decay particle overlaps spatially with prompt energy due to other particles produced in the neutrino interaction. Because of this effect, the reconstruction efficiency depends on the particle multiplicity.

The timing resolution is a function of the number of observed photoelectrons (PE) because it is based on the decay of the fluors in the scintillator and wavelength-shifting fiber. For many-PE hits, the timestamp will come from light that resulted from very prompt decays of the scintillator and fiber; at smaller numbers of PE, the recorded hit times are delayed relative to the true time of the energy deposition. The timing resolution is 10~ns for 1-2 PE hits, 3~ns for 6-12 PE hits due to minimum ionizing particles, and approaches the 2.2~ns resolution of the electronics asymptotically at very high pulse heights.

Timing information is first used to separate multiple neutrino interactions within a single 10 $\mu$s beam pulse. Hits are sorted by their time, and a scan is performed to find 80-ns windows where the total energy exceeds a threshold. The window is moved forward in time until the threshold is no longer met. This algorithm reliably separates neutrino interactions which occur 100~ns apart, and keeps a \kp and its decay products together. 

The design, calibration and performance of the \minerva detector, including the calibration of the timing response, is described in detail in Ref.~\cite{minerva_nim}. The hit timing in the data acquisition system is described in Ref.~\cite{daq_nim}. The data for this measurement were collected between March 2010 and April 2012, corresponding to $3.51\times 10^{20}$ protons on target (POT). The horn current was configured to focus positive pions, resulting in a \numu-enriched beam with 10\% \numubar contamination which is largely in the high-energy tail of the flux.

\section{Experiment Simulation}
\label{sec:expt_sim}
The neutrino beam is simulated by a Geant4-based model~\cite{geant4a, geant4b} that is tuned to agree with hadron production measurements on carbon~\cite{na49, na49b} by the procedure described in Ref.~\cite{MINERvAflux,leothesis}. Uncertainties on the neutrino flux arise from the statistical and systematic uncertainties in these hadron production experiments, as well as uncertainties in the beamline geometry and alignment~\cite{zarko}. The integrated neutrino flux is estimated to be $\unit[(2.95 \pm 0.23) \times 10^{-8}]{cm^{-2}/POT}$. Table~\ref{tab:nu_flux} lists the flux as a function of energy.

\begingroup
\squeezetable
\begin{table*}[t]
\centering
\begin{tabular}{l|ccccccccccccc}
\hline \hline
$E_\nu$ (GeV) & 
$0 - 1$ &
$1 - 2$ &
$2 - 3$ &
$3 - 4$ &
$4 - 5$ &
$5 - 6$ &
$6 - 7$ &
$7 - 8$ \\
Flux ($10^{-9}~\nu_{\mu}$/cm$^2$/POT) &
$1.0331  $ &
$4.3611  $ &
$7.4333  $ &
$7.9013  $ &
$3.2984  $ &
$1.2193  $ &
$0.7644  $ &
$0.5671  $ \\
\hline
$E_\nu$ (GeV) & 
$8 - 9$ &
$9 - 10$ &
$10 - 15$ &
$15 - 20$ &
$20 - 25$ &
$25 - 30$ &
$30 - 35$ &
$35 - 40$ \\
Flux ($10^{-9}~\nu_{\mu}$/cm$^2$/POT) &
$0.4398  $ &
$0.3834  $ &
$1.1135  $ &
$0.4664  $ &
$0.1959  $ &
$0.0918  $ &
$0.0668  $ &
$0.0553  $ \\
\hline
$E_\nu$ (GeV) & 
$40 - 45$ &
$45 - 50$ &
$50 - 60$ &
$60 - 70$ &
$70 - 80$ &
$80 - 90$ &
$90 - 100$ &
$100 - 120$ \\
Flux ($10^{-9}~\nu_{\mu}$/cm$^2$/POT) &
$0.0474  $ &
$0.0325  $ &
$0.0267  $ &
$0.0057  $ &
$0.0023  $ &
$0.0006  $ &
$0.0001  $ &
$0.0000  $ \\
\hline
\hline
\end{tabular}
\caption{The predicted $\nu_\mu$ flux per POT for the data included in this analysis.}
\label{tab:nu_flux}
\end{table*}
\endgroup

Neutrino interactions are simulated using the \genie 2.8.4 neutrino event generator~\cite{genieNIM}. Kaons are produced via the decays of baryon resonances as well as from hadronization in DIS events. In \genie, individual resonances are simulated only for hadronic invariant masses $W < 1.7$~GeV, which is just above the $K^{+} \Lambda$ threshold, and most \kp originate in hadronization. A parameterization based on Koba-Nielsen-Olese (KNO) scaling~\cite{kno} is used for $1.7 < W < 2.3$~GeV and PYTHIA6~\cite{pythia} is used for $W > 3.0$~GeV. In the intermediate region $2.3 < W < 3.0$~GeV, the AGKY model~\cite{agky} governs the transition between KNO and PYTHIA6. Parameters which control the rate of strange particle production in hadronization are tuned such that rates of \lam and \kzS production on deuterium and nuclei agree with BEBC~\cite{bebc1, bebc2, bebc3, bebc4} and Fermilab 15'~\cite{fnal15a, fnal15b} bubble chamber measurements as a function of $W$. All events in \genie 2.8.4 are $\Delta S = 0$, so there is always another strange particle in the final state in addition to the $K^{+}$.

Final-state interaction processes are simulated using an effective intranuclear cascade called the ``hA'' model~\cite{dytmanFSI}, which simulates the full cascade as a single interaction and tunes the overall interaction rate to hadron-nucleus total reaction cross section data. Kaon rescattering was added to \genie in version 2.8.0, and is tuned to data from Bugg \textit{et al.}~\cite{bugg} and Friedman \textit{et al.}~\cite{friedman}. Charge exchange processes are included for pions but not kaons. Because \kp cannot be absorbed due to strangeness conservation, and \genie does not simulate \kp production in pion reactions, the \genie FSI model never adds or removes \kp from the final state. Rescattering of the \kp occurs in 13\% of simulated events in this analysis sample, reducing the final-state \kp kinetic energy. 

The \minerva detector response is simulated by a Geant4-based model using Geant4 version 9.4.p02. Interactions of \kp in the detector affect the range-based measurement of the \kp kinetic energy. The version of Geant4 used in this analysis has a wiggle in the \kp-carbon inelastic cross section at low \kp energy, and does not agree with external data. This feature is also observed by the T2K collaboration with version 9.4.4~\cite{martti} but is not present in the newer version 10.0~\cite{hans}. To correct for this feature, a weight is applied to simulated events based on whether the \kp scatters inelastically, elastically, or not at all. The weights are given in Eq.~\ref{eq:kcwgts}:

\begin{equation}
 \label{eq:kcwgts}
  \begin{split}
    W_{inel} & = \frac{1-e^{-\rho x \sigma_{data}^{tot}}}{1-e^{-\rho x\sigma_{geant}^{tot}}} \times \frac{\sigma_{data}^{inel}}{\sigma_{data}^{tot}} \times \frac{\sigma_{geant}^{tot}}{\sigma_{geant}^{inel}} \\
    W_{el} & = \frac{1-e^{-\rho x \sigma_{data}^{tot}}}{1-e^{-\rho x\sigma_{geant}^{tot}}} \times \frac{\sigma_{data}^{el}}{\sigma_{data}^{tot}} \times \frac{\sigma_{geant}^{tot}}{\sigma_{geant}^{el}} \\
    W_{none} & = e^{-\rho x(\sigma_{data}^{tot}-\sigma_{geant}^{tot})},
  \end{split}
\end{equation}

\noindent
where $\rho$ is the density of the tracker and $x$ is the distance traveled by the $K^{+}$. The Geant4 prediction $\sigma_{geant}^{el}$ ($\sigma_{geant}^{inel}$) is taken from a spline fit to cross sections determined by counting elastic (inelastic) interactions in a simulation of \kp incident on a thin carbon target. The total cross section prediction $\sigma_{geant}^{tot}$ is the sum of the elastic and inelastic components. The data constraint $\sigma_{data}^{tot}$ is a parameterization of \kp-carbon total cross section measurements~\cite{bugg,friedman}. The inelastic constraint $\sigma_{data}^{inel}$ is a parameterization of reaction cross section data~\cite{friedman}, and includes nucleon knock-out. The elastic component $\sigma_{data}^{el}$  is not measured directly. Its shape as a function of \kp energy is taken from Geant4, and it is normalized to agree with the average difference between the total and inelastic data. Scattering on other nuclei in the tracker is reweighted based on the carbon data and $A$-dependent nuclear effects are not considered.

\begin{figure}[h]
\centering
\includegraphics[width=\columnwidth]{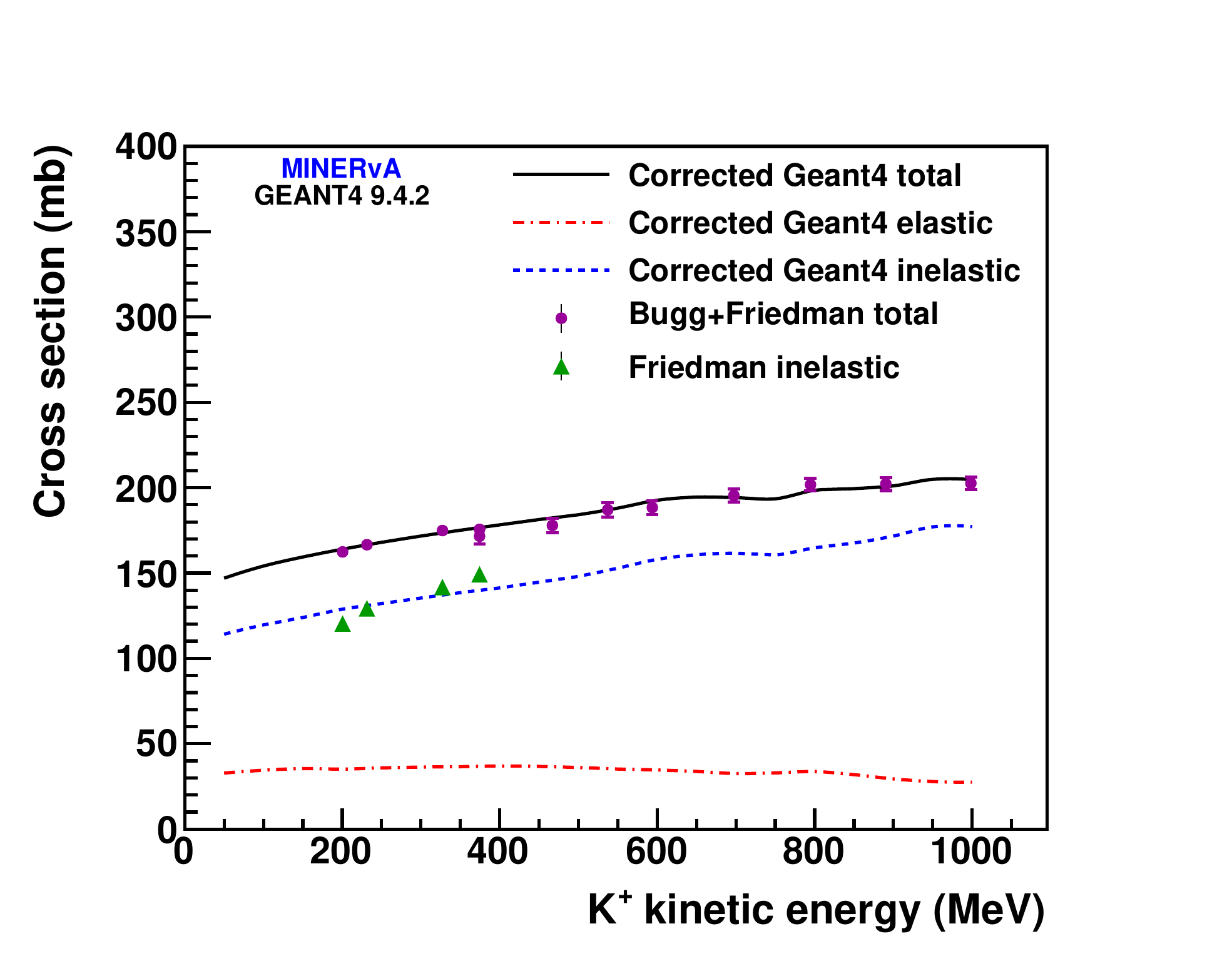}
    \vspace{-7pt}
\caption{(color online) Kaon interactions predicted by Geant4 are reweighted to agree with external data from Bugg \textit{et al.}~\cite{bugg} and Friedman \textit{et al.}~\cite{friedman} on the total and inelastic scattering cross sections on carbon. \label{fig:kaon_carbon_xs}}
    \vspace{-10pt}
\end{figure}

For events with \kp kinetic energy less than $600$~MeV, 16\% undergo only elastic scattering (reduced to 10\% by reweighting), 28\% experience inelastic reactions (increased to 34\%), and the remaining 56\% of events have no \kp interaction. After reweighting, the inelastic and total scattering cross sections as a function of kaon energy agree with external \kp-carbon scattering data, as shown in Fig.~\ref{fig:kaon_carbon_xs}. 

Because of the energy dependence of the probability for a reaction to occur as the \kp propagates through the detector, the true \kp kinetic energy spectrum in the simulation used for this analysis is 12\% below \genie's untuned prediction at kaon energies up to about 400 MeV, rising to 4\% above \genie at kaon energies greater than 500 MeV. The effect on the extracted cross section of warping the true \kp kinetic energy spectrum in this way was studied and found to be negligibly small.

Through-going muons originating from neutrino interactions in the rock upstream of \minerva are used to calibrate the detector energy scale for individual hits. The energy scale is determined to within 2\% by requiring that the PE and reconstructed energy be the same in data and simulation. The hit energy is further corrected for passive materials in the tracker and calorimeters. Measurements made with a miniature version of the MINERvA detector in a hadron test beam are used to determine Birks' parameter for the scintillator, as well as the energy response to single pions and protons~\cite{testbeam}.

\begin{figure}[h]
\centering
\includegraphics[width=\columnwidth]{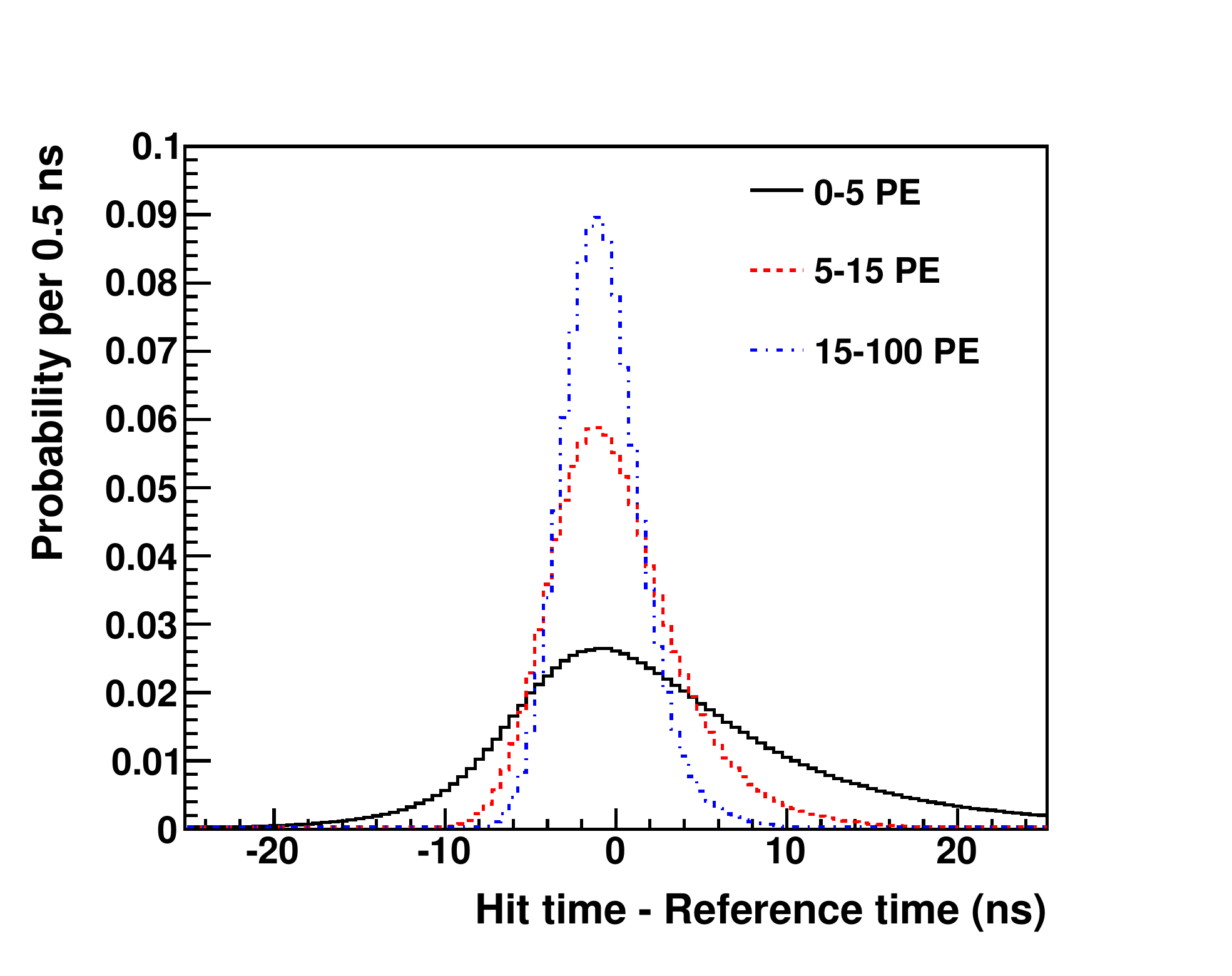}
    \vspace{-7pt}
\caption{(color online) The probability density functions used in the simulation and for the timing fit are constructed using through-going muons in data. Three bins of PE are shown here as an illustration; in the actual fit, finer bins are used. \label{fig:timingpdfs}}
    \vspace{-10pt}
\end{figure}

Hit times are simulated by smearing the true time of an energy deposit using a function derived from data. The function is determined by taking fully-calibrated throughgoing muon tracks and comparing the reconstructed hit times to the track ``reference time'' in slices of the number of PE. The reference time is determined by fitting the timing profile of hundreds of hits along the track, and is known to within 1~ns. Corrections are applied to the hit times to account for muon time-of-flight, light time-of-flight in the optical fiber, timing offsets inherent to the electronics, and the expected delay between the energy deposit and the earliest decay of the scintillator and fiber. The probability density functions for hits in three different pulse height bins are shown in Fig.~\ref{fig:timingpdfs}. The light time-of-flight correction assumes that the earliest photoelectron is the result of light that traveled directly from the muon energy deposit to the PMT. Especially for low pulse height hits, it is possible that all photoelectrons result from light that first reflected off the mirrored end of the optical fiber. This gives rise to a high-side tail in Fig.~\ref{fig:timingpdfs} that reduces with increasing PE.

Pile-up due to multiple neutrino interactions within a single beam 10-$\mu$s beam pulse is an important background in this analysis. Simulated events are generated one per beam pulse, then overlaid on top of a pulse taken from data. The algorithm which separates multiple neutrino interactions based on timing is run on the entire collection of hits. When the simulated event occurs close in time to a data event, the two are reconstructed together. The majority of the detector activity responsible for this pile-up is due to neutrino interactions in either the rock upstream of \minerva or in the side hadronic calorimeter.

\section{Event Reconstruction}
\label{sec:reco}

We define the signal process as a \numu-induced charged-current reaction with at least one \kp exiting the nucleus in which the neutrino interaction occurred. The fiducial volume is a 5.57-ton region in the \minerva tracker, beginning 18~cm downstream of the last passive nuclear target, ending 25~cm upstream of the rear electromagnetic calorimeter, and with a hexagonal apothem 21~cm smaller than that of the inner detector.

Kaons are selected by reconstructing the timing signature of a \kp decay-at-rest. This requires that the \kp stop inside the tracker or electromagnetic calorimeter. If the \kp stops in the hadronic calorimeter, 90\% of the energy from its decay products is deposited in passive material and the \kp cannot be reliably reconstructed. Non-interacting kaons with more than 600 MeV of kinetic energy typically reach the hadronic calorimeter and cannot be reconstructed using this timing-based technique. High-energy kaons are reconstructed only when they interact inelastically inside the tracker, in which case the range-based kinetic energy measurement is poor. The differential cross section will be presented from 0 to 500 MeV of \kp kinetic energy.

The timing signature reconstruction begins with a search for activity in the detector that is delayed in time with respect to the neutrino interaction, consistent with the 12.4~ns \kp lifetime, and consistent in energy with the products of a \kp decay-at-rest. First, we search for a fully-reconstructed $K^{+} \rightarrow \mu^{+}$ decay. Using only topological information, we find tracks which kink, using an algorithm described in Ref.~\cite{minerva_nim}.

If no kinked track is found, hits not associated with any tracks are grouped into narrow bunches in time, called ``time slivers,'' with a granularity of 5~ns. Events are accepted if there is a delayed time sliver that is spatially near the endpoint of a \kp track. We also search for delayed time slivers near the neutrino interaction vertex. This extends the acceptance to \kp with very small kinetic energy. These two samples are combined with additional selections to purify the \kp content. These selections are summarized in Table~\ref{tab:cuts} and described in detail below.

\begingroup
\squeezetable
\begin{table*}[t]
\begin{tabular}{cccccccccc}
\hline\hline
Cut & Data & MC Total & Efficiency (\%) & Purity (\%) \\
\hline
Kink likelihood ratio & 500 & 512 & 1.9 & 50.2 \\
Kink secondary energy & 394 & 424 & 1.8 & 57.7 \\
\hline
Decay sliver time gap & 35577 & 36590 & 13.4 & 4.9 \\
Decay sliver energy & 7503 & 7698 & 9.2 & 15.8 \\
Decay sliver number of hits & 3826 & 3561 & 7.5 & 28.1 \\ 
Distance to decay sliver & 2369 & 2372 & 7.1 & 39.6 \\
\hline\hline
Kaon by any method & 2763 & 2796 & 8.9 & 42.3 \\
Longest track range & 2155 & 2198 & 8.1 & 48.7 \\
Non-kaon hadronic visible energy & 1837 & 1878 & 7.5 & 53.3 \\
Low-energy event scan & 1688 & 1700 & 7.3 & 56.8 \\
\hline\hline
\end{tabular}
\caption{A summary of selected events, efficiency and purity after each cut for kaons below 500 MeV of kinetic energy. The numbers shown are cumulative. The kinked track (top section) and decay bunch (middle section) selections are combined to form the final sample (bottom section). }
\label{tab:cuts}
\end{table*}
\endgroup

\begin{figure*}[t]
\centering
\ifnum\DoPrePrint=1
  \includegraphics[width=\columnwidth]{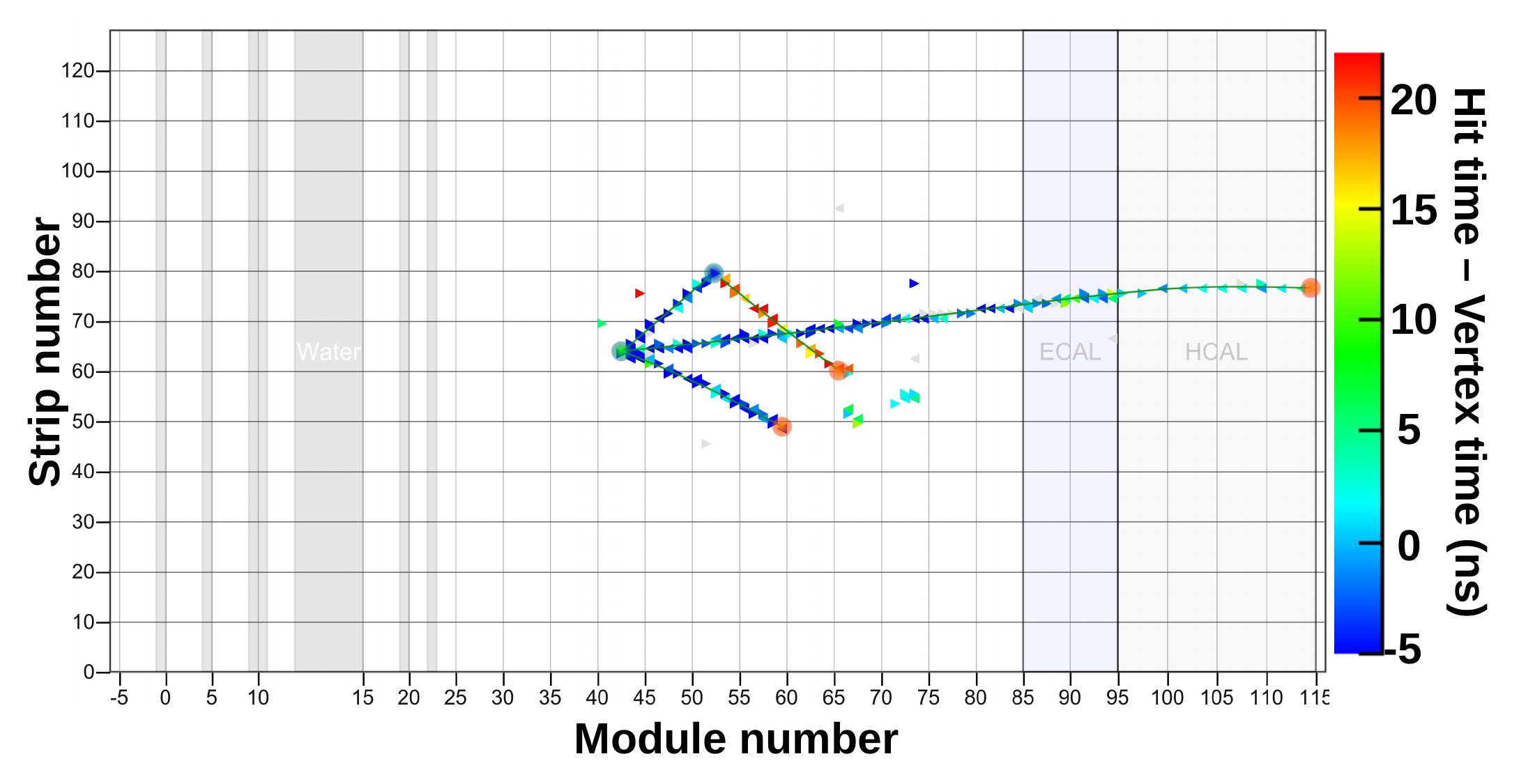}
\else 
  \includegraphics[width=1.8\columnwidth]{kaon_event}
\fi
    \vspace{-7pt}
\caption{(color online) A \numu-induced charged-current \kp candidate in MINERvA data is viewed from above. The beam is angled into the page at $3.5^{\circ}$ with respect to the horizontal axis. Each colored triangle represents one hit, a time-stamped energy deposit in a single scintillator strip. The color represents the hit time, relative to the reconstructed time of the interaction. The green circle is the event vertex, orange circles are reconstructed track endpoints, and the blue circle is a track kink. The green lines are reconstructed tracks. The kinked track is the \kp candidate. The longest track is the muon candidate and is matched to a negatively-charged track in MINOS. The second segment of the kinked track is a \mup from the decay-at-rest $K^{+} \rightarrow \mu^{+} \nu_{\mu}$, with a time gap between the two segments of 18 ns. The remaining particles are likely the decay products of $\Sigma^{+} \rightarrow \pi^{+} n$, where the \pip is the other track and the detached hits are proton products of a scattering neutron. \label{fig:kaon_event}}
    \vspace{-10pt}
\end{figure*} 

In kinked track events, hit times are corrected for time-of-flight and fit under two hypotheses. In the first hypothesis, the two segments are assumed to have the same true time, as would be the case for a pion that undergoes a hard scatter. In the second hypothesis, the true times of the two segments are allowed to float independently. For signal events, the second segment is due to the \mup from \kp decay, and will be late in time relative to the first segment, which is the \kp track.

The probability density functions used in the timing fit are identical to those used in the simulation and examples are shown in Fig.~\ref{fig:timingpdfs}. The fit maximizes the sum over all hits of the natural logarithm of the probability density. By construction, the two-parameter kaon decay hypothesis always gives a better fit, and the value of the log-likelihood ratio is zero when the best-fit time gap is zero. 

\begin{figure*}[t]
\centering
\ifnum\DoPrePrint=1
  \subfloat[] {
    \includegraphics[width=0.5\columnwidth]{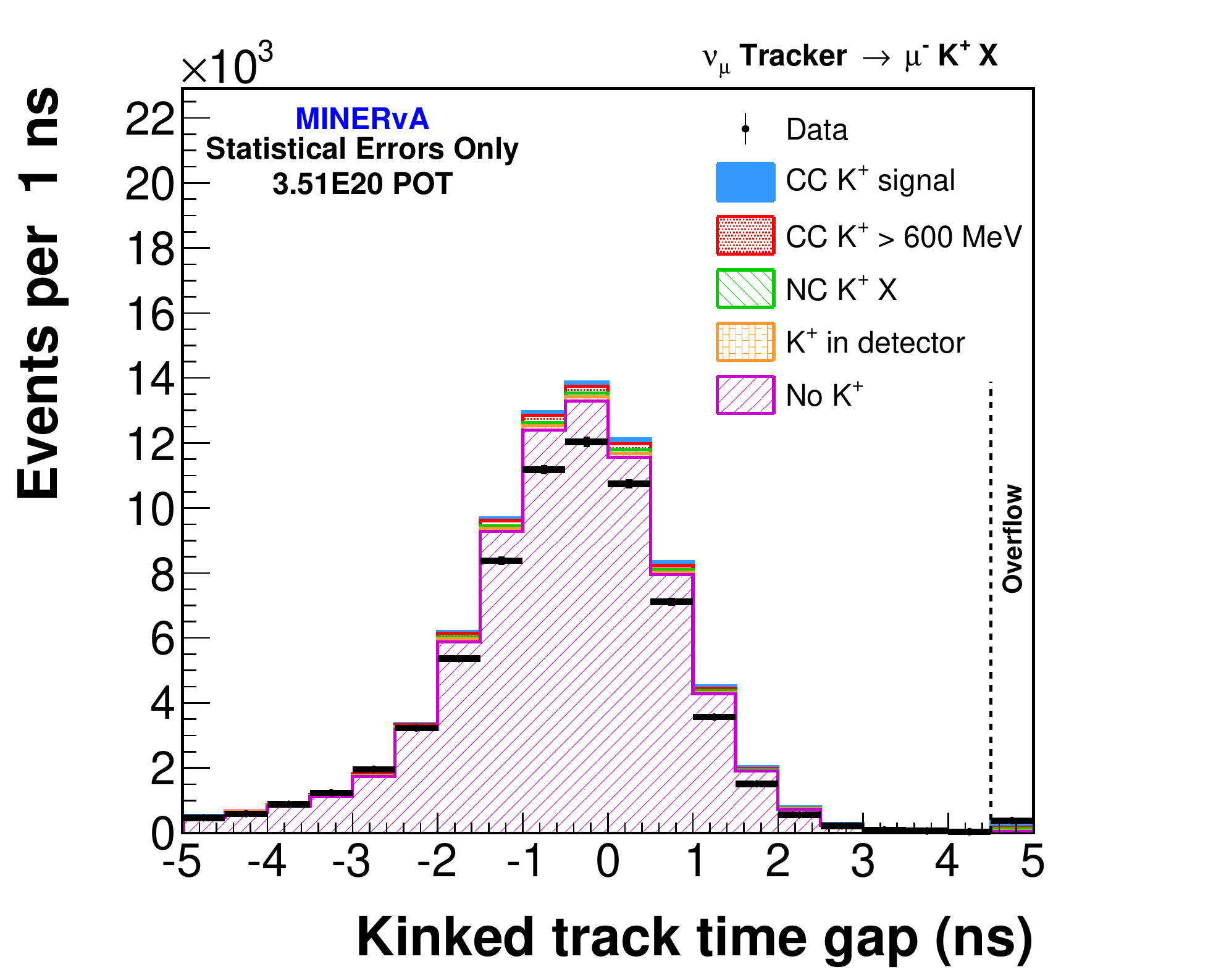}
  }
  \subfloat[] {
    \includegraphics[width=0.5\columnwidth]{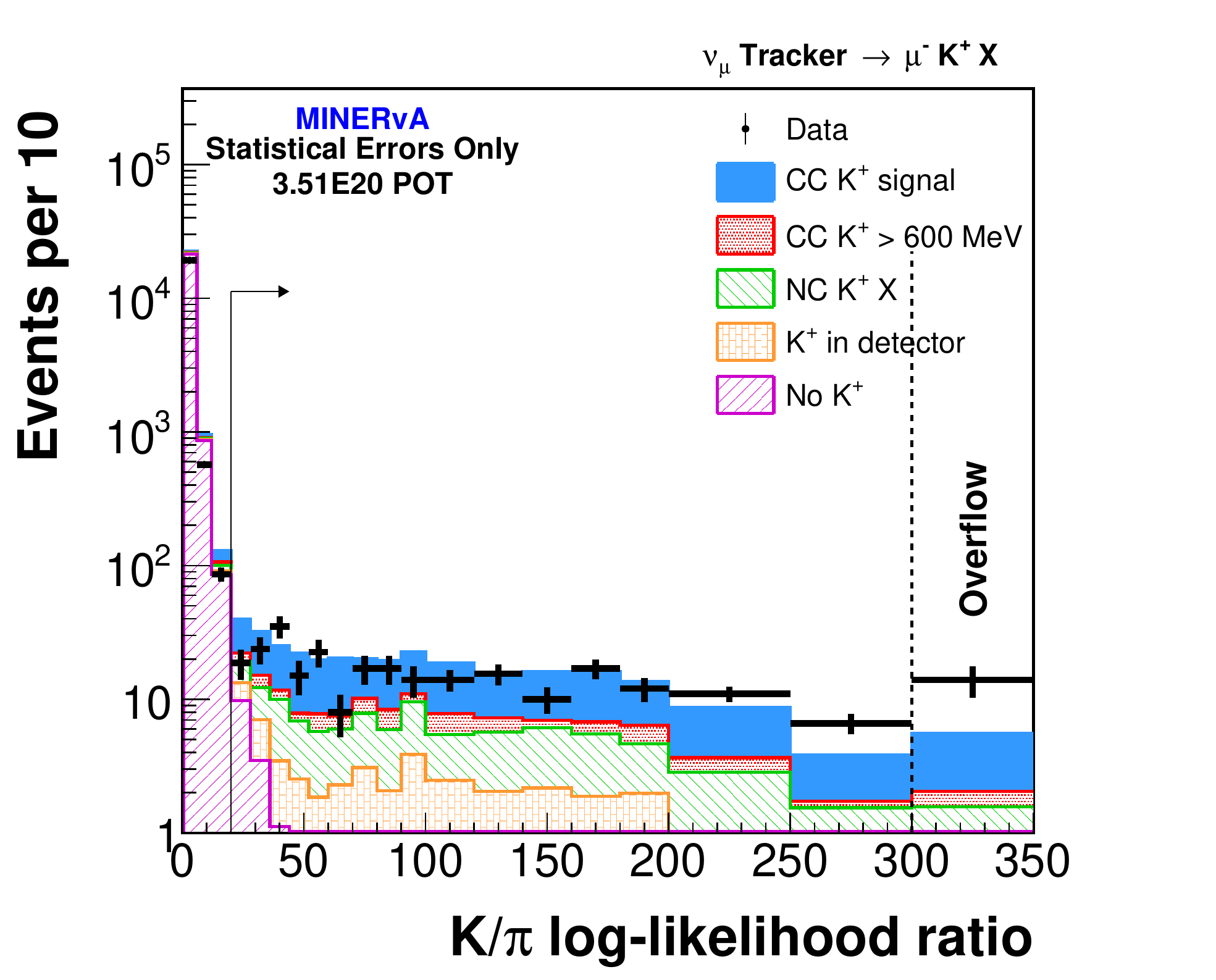}
  }
\else 
  \subfloat[] {
    \includegraphics[width=\columnwidth]{kink_time_peak}
  }
  \subfloat[] {
   \includegraphics[width=\columnwidth]{kink_K_pi_likelihood}
  }
\fi
\caption{(color online) The time gap between primary and secondary segments of a kinked track (left) agrees in shape in data and simulation. The peak region is mostly due to interacting pions, where the deficit in data relative to simulation is consistent with other results indicating an overprediction in \genie's pion production model~\cite{minerva_pion}. The log-likelihood ratio of the fit described in the text separates stopping kaons from interacting hadrons (right). The arrow shows the selection of events with log-likelihood ratio greater than 20. \label{fig:kinktracks}}
\end{figure*}

An example signal candidate data event is shown in Fig.~\ref{fig:kaon_event}. The time gap distribution for a background-rich sample and the log-likelihood ratio distribution are shown in Fig.~\ref{fig:kinktracks}. The peak in the time gap plot is slightly below zero because the time-of-flight correction assumes a low-energy stopping kaon. The majority of the background events with small time gap come from interacting pions, which essentially travel at the speed of light. The low-side tail is due to events where the track direction is truly backward but is reconstructed as forward. In these events, the first and second segments are reversed, and the time-of-flight correction goes in the wrong direction.

If no tracks are found to have kinks, we consider time slivers of untracked hits. Typical out-of-time energy deposits can be separated if they are more then 10 ns apart. A sliver is considered a \kp decay product candidate if its best-fit time is at least 9 ns later than the time of the stopping \kp track candidate. Slivers with small time gaps relative to the track are typically due to interacting pions or protons, or due to other activity from the primary neutrino interaction. About half of true stopping kaons are rejected because the decay occurs promptly and cannot be separated from the much larger background from interacting hadrons.

Events can be accepted even in the absence of a \kp track. When no stopping track is found, delayed time slivers are considered kaon decay candidates if they are at least 11 ns later than the time of the muon track, and spatially near the neutrino interaction point. This extends the acceptance to \kp kinetic energies below the tracking threshold, which is approximately 100 MeV. Events selected by this method are scanned using the Arachne event visualization program~\cite{arachne}, and a straight line is drawn by eye connecting the start of the muon track to the nearest delayed energy deposit. The \kp kinetic energy is estimated from the length of the line segment measured in $g/cm^{2}$ based on the simulation.

\begin{figure*}[t]
\centering
\ifnum\DoPrePrint=1
  \subfloat[] {
    \includegraphics[width=0.5\columnwidth]{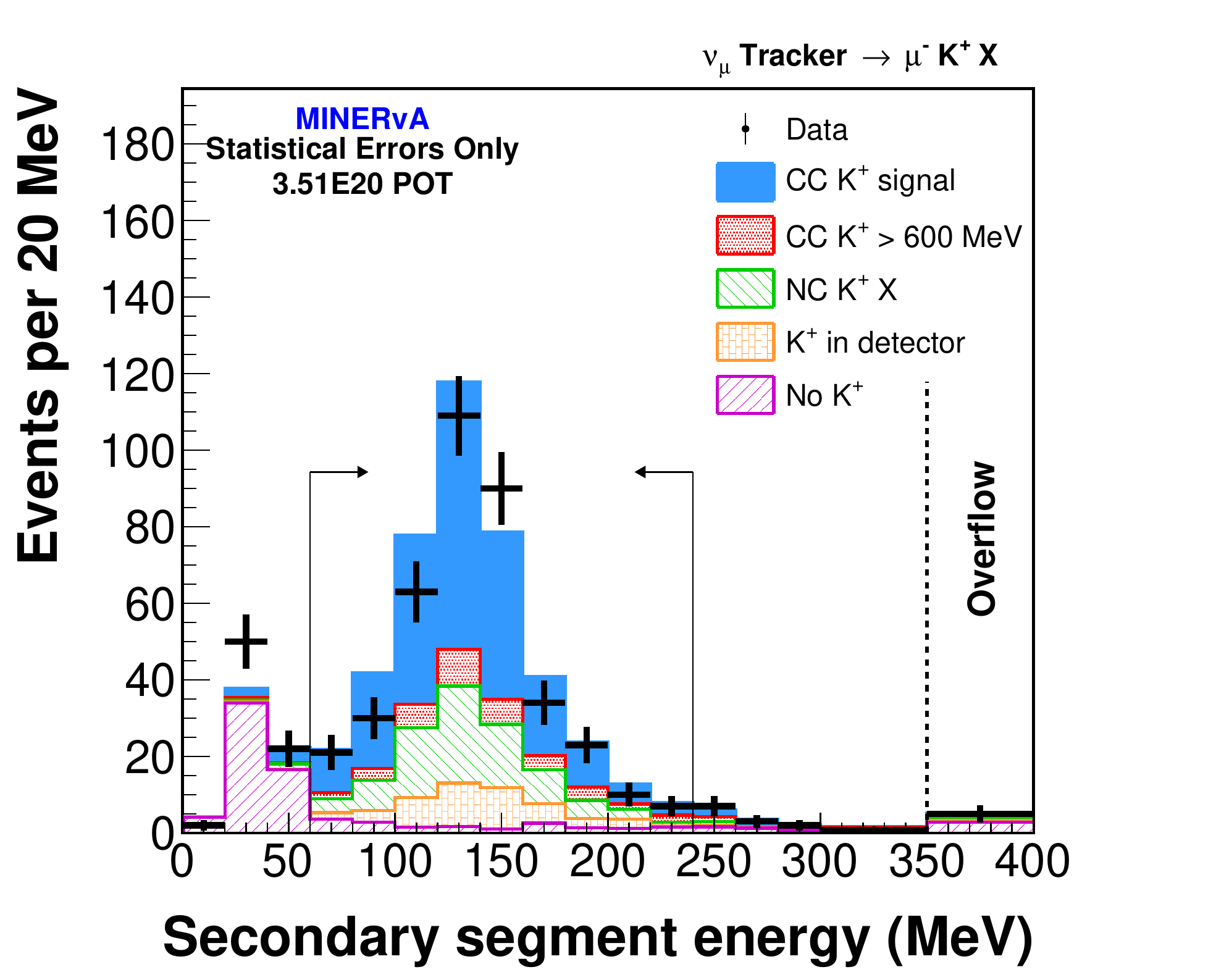}
  }
  \subfloat[] {
    \includegraphics[width=0.5\columnwidth]{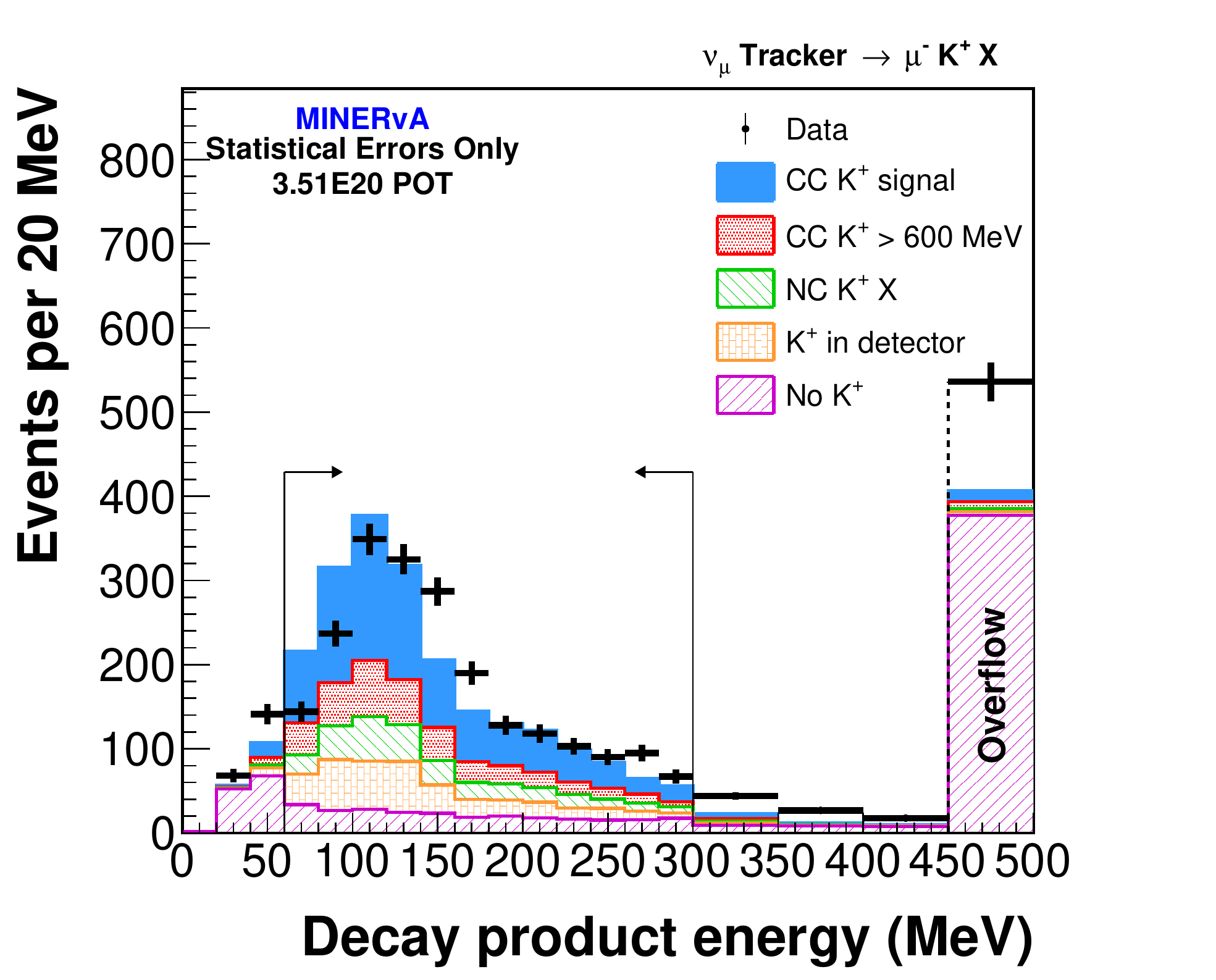}
  }
\else 
  \subfloat[] {
    \includegraphics[width=\columnwidth]{kink_secondary_energy}
  }
  \subfloat[] {
   \includegraphics[width=\columnwidth]{decay_product_energy}
  }
\fi
\caption{(color online) The energy of the \kp decay product candidate when it is tracked (left) and untracked (right). Below 60 MeV, the largest background in both cases is due to Michel electrons. At high energies in the right (untracked) plot, the background is predominantly due to pile-up, which is undersimulated by 21\%. The arrows show the selection criteria. \label{fig:decayenergy}}
\end{figure*}

Multiple independent visual scans were carried out on a sample of data and simulated events mixed together randomly, such that the scanner had no knowledge of whether a given event came from data or simulation. A control sample of 500 events was scanned by all scanners and used to estimate the level of disagreement between scanners in the amount of material the \kp passed through. The average disagreement over the 500-event sample corresponds to 18 MeV of kaon kinetic energy. Potential systematic biases between individual scanners were studied and determined to be much smaller than the 18 MeV average disagreement. An uncertainty is added to account for differing results. For range-out kaons, the full width at half-maximum of the kinetic energy residual is approximately 20 MeV for tracked kaons, and 35 MeV for kaons whose energy is measured by the scan.

In 20\% of the visually scanned events, there are no hits due to a charged particle connecting the neutrino interaction vertex and the nearest delayed hits. No kinetic energy can be estimated in these events and they are rejected. Of the rejected events in simulation, 45\% are due to pile-up, in which the delayed time sliver is due to a subsequent neutrino interaction. In total, 46\% of background events and only 9\% of true signal events are rejected. Signal events are typically rejected when the kaon decay is $\kp \rightarrow \pip \piz$ and the \pip is obscured by prompt hits. The two \piz photons are reconstructed, but because of the gap between the \piz decay and photon conversion, the point where the \kp stopped cannot be determined. 

Additional selection cuts are applied in order to reject events where the delayed energy is actually due to a ``Michel'' electron from the decay chain $\pi \rightarrow \mu \rightarrow e$. A \kp at rest will decay to a \mup with 152 MeV of kinetic energy (and an unobserved neutrino) 64\% of the time, and a back-to-back \pip and \piz 20\% of the time. Both of these decay modes, as well as other less probable decays such as $e^{+} \pi^{0}$ will deposit approximately 150 MeV of energy in the \minerva detector. The endpoint of the Michel electron spectrum is 55 MeV, and we select events with at least 60 MeV of reconstructed energy. 

The distribution of decay product energy is shown in Fig.~\ref{fig:decayenergy} for kinked track and delayed time sliver events separately. The energy in these plots is delayed by $5-60$ ns relative to the \kp, but does not include a Michel electron from $\kp \rightarrow \mup \rightarrow e^+$, which is observed much later in time. Energy deposits due to \kp decay products are not included when they occur on scintillator strips that are also intersected by the \kp itself, as the hit timestamp comes from the earliest energy. This reduces the peak observed energy deposit from the \kp decay products. The distribution for delayed time sliver events is wider because of the contribution from decay modes other than $\kp \rightarrow \mup \numu$. The visible energy is greatest for $\kp \rightarrow e^+ \piz$.

In events that do not have fully-reconstructed kinked tracks, we require that the delayed time sliver have hits in at least 10 different strips. This requirement rejects events due to neutrons, which can scatter in the detector to produce low-energy knock-out protons late in time. These events typically produce large energy deposits in a small number of strips.

Pile-up from multiple neutrino interactions in the same 10 $\mu$s beam pulse can fake the timing signature of a \kp decay at rest. To reduce this background, we require the mean distance from the kaon endpoint vertex to a hit in the delayed time sliver to be less than 80 cm. The largest contribution is due to neutrino interactions in the side hadronic calorimeter that leak energy into the inner detector.

Charged-current events are selected by requiring that a track other than the \kp candidate traverse more than 250 $g/cm^{2}$ of material in \minerva, where the side and downstream calorimeters are included. Events with muons below 500 MeV of kinetic energy are rejected. For muons that are matched with tracks in MINOS (42\% of the sample), we require the curvature to be consistent with a negatively-charged particle to remove antineutrino-induced events. In the simulation, 3.9\% of muons that are not matched into \minos are \mup from antineutrino events, and are subtracted.

Hadronic interactions of high-energy charged pions can produce $K^{+}$, for example $\pi^{+} n \rightarrow K^{+} \Lambda$. The \kp can then stop and decay in the detector and mimic the signal. When this process takes place inside the nucleus of the neutrino interaction, the event is considered signal. However, when it occurs elsewhere in the detector, it must be subtracted. These events produce large hadronic showers, with an average pion energy of 3.3 GeV according to the simulation. To remove these events, we sum the hadronic energy in the detector, excluding the \kp track. This energy includes the particles produced in the neutrino interaction, as well as products of their subsequent hadronic interactions. 31\% of such events are rejected by requiring that the calorimetrically-corrected hadronic energy be less than 8 GeV. 

High-energy kaons which interact hadronically inside the detector are misreconstructed at much lower kinetic energy. As in the case of $\pip \rightarrow \kp$ interactions, high-energy hadronic showers are produced, and 24\% of interacting \kp with true kinetic energy \textgreater~600 MeV are rejected by the cut on non-\kp hadronic energy, which includes the products of the \kp interaction. A sideband of events with non-kaon hadronic energy \textgreater~8 GeV is used to constrain these two classes of events simultaneously.

\begin{figure}[h]
\centering
\includegraphics[width=\columnwidth]{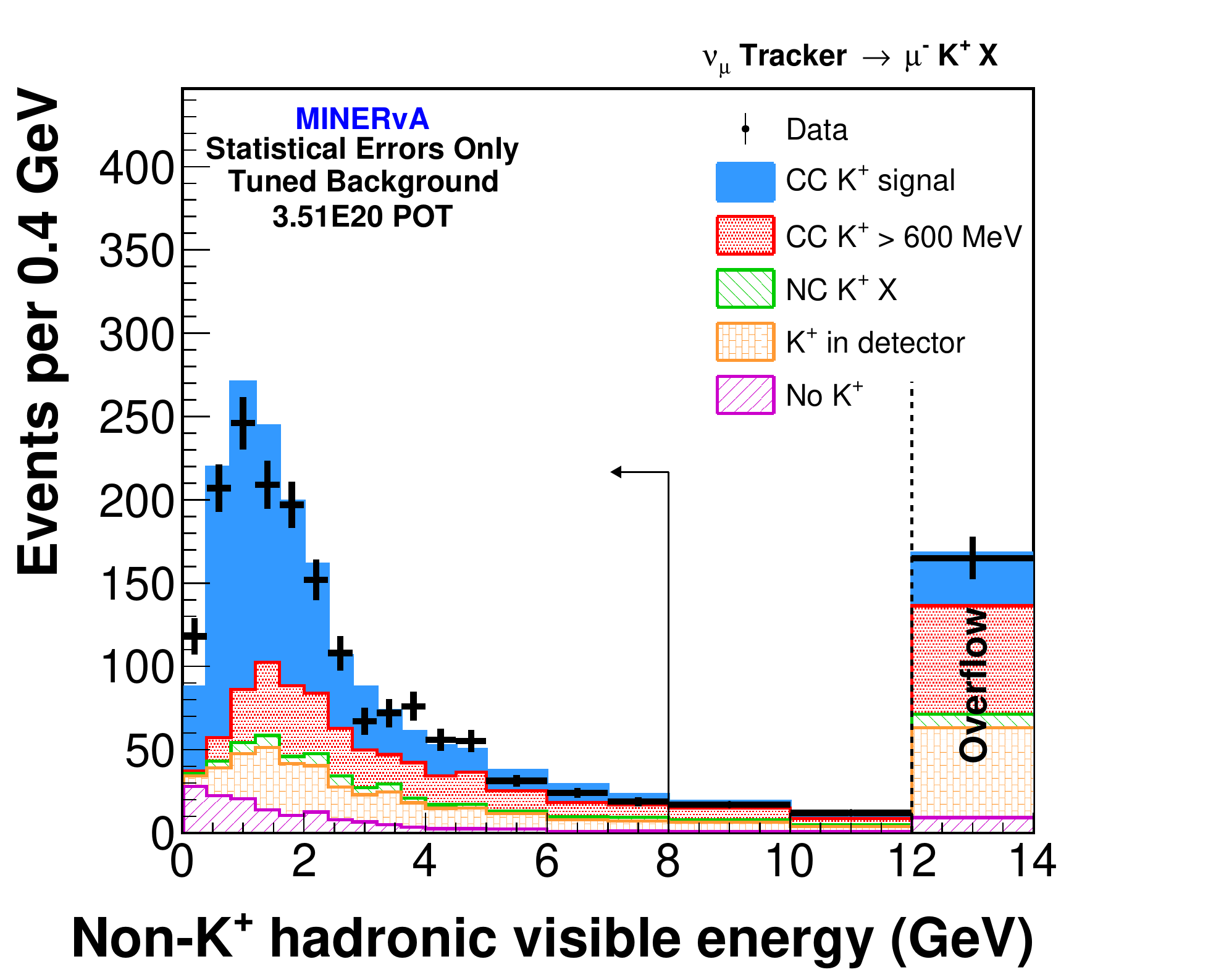}
    \vspace{-7pt}
\caption{(color online) The reconstructed non-$K^{+}$ hadronic visible energy is used to reject and constrain events with high-energy hadronic interactions in the detector but outside the nucleus of the neutrino interaction. The arrow shows the cut at 8 GeV; events to the right of the cut are used in the sideband constraint.\label{fig:numinusek}}
    \vspace{-10pt}
\end{figure}

The distribution of reconstructed non-kaon hadronic energy is shown in Fig. \ref{fig:numinusek}. The highest bin is overflow and is not bin-width normalized. After all cuts are applied, 1755 events are selected in data prior to background subtraction. A summary of event selection cuts is given in Table \ref{tab:cuts}.

\section{Cross Section Extraction}

We report a differential cross section with respect to the \kp kinetic energy, $T_K$. Neutrino energy, muon energy, and muon angle are not reported due to the limited kinematic region where muon energy can be measured. Energy cannot be measured for muons with energy greater than 1 GeV and angle wider than 17$^\circ$ (47\% of signal events), as they exit \minerva and are not reconstructed in \minos. Kaon angle is not reported due to the difficulty in measuring the angle of \kp at low energies, where the effect of FSI is expected to be the largest. The flux-integrated differential cross section per nucleon in bin $i$ is

\begin{equation} \label{eq:diffxs}
\left(\frac{d\sigma}{dT_{K}}\right)_{i} = \frac{\sum_{j}U_{ij}\left(N_{j} - N_{j}^{bg}\right)}{\epsilon_{i}N_{nuc}\Phi\Delta_{i}},
\end{equation}
where $j$ is the index of a reconstructed $T_{K}$ bin, $U_{ij}$ is the unsmearing matrix, $N_{j}$ is the number of selected events, $N_{j}^{bg}$ is the predicted number of background events, $\epsilon_{i}$ is the selection efficiency for signal events, $N_{nuc}$ is the number of nucleons in the fiducial volume, $\Phi$ is the integrated \numu flux prediction, and $\Delta_{i}$ is the width of bin $i$.

\subsection{Background Subtraction}

\begingroup
\squeezetable
\begin{table}
\begin{tabular}{ccc}
\hline\hline
Category & Signal region (\%) & sideband region (\%) \\
\hline
CC \kp, $T_{K}$ \textless~ 600 MeV & 54.2 & 22.5 \\
CC \kp, $T_{K}$ \textgreater~ 600 MeV & 19.5 & 36.6 \\
\pip $\rightarrow$ \kp & 5.8 & 15.7 \\
\kz $\rightarrow$ \kp & 5.3 & 10.0 \\
$\bar{\nu}_{\mu}$-induced or outside F.V. & 4.6 & 3.9 \\
NC \kp & 3.6 & 6.2 \\
Pile-up & 4.8 & 3.9 \\
Other & 2.4 & 1.1 \\
\hline\hline
\end{tabular}
\caption{The breakdown of selected events for the signal and high hadronic energy sideband regions in the simulation, expressed as a percentage of the total samples prior to sideband tuning. ``Other'' includes events which are truly due to slow neutrons or Michel electrons.}
\label{tab:bkg}
\end{table}
\endgroup

The predicted background from simulation is scaled to agree with data in two sideband regions. The background due to beam pile-up is constrained by events where the mean distance from the kaon endpoint to a hit in the delayed time sliver is greater than 120 cm. In this region, 86\% of the events are due to pile-up. A fit is performed to determine the scale factor for the pile-up events, with other classes of events held fixed. The extracted scale factor of 1.21 is applied to backgrounds caused by pile-up.

Backgrounds from \kp production by \pip reactions in the detector are constrained along with kaons with true kinetic energy \textgreater~600 MeV from a sideband of events with non-kaon hadronic energy greater than 8 GeV. A single scale factor of 1.08 is determined and applied to all backgrounds, except to those due to beam pile-up.

Signal events with true \kp kinetic energy \textless~600 MeV comprise 22.5\% of the sideband region. This introduces a small uncertainty due to the signal normalization into the analysis. A cross section is initially extracted leaving the normalization of these events fixed. A scale factor of 0.90 $\pm$ 0.13 is computed from the ratio of the integrated cross section in data and simulation. The analysis is repeated by applying that scale factor and its associated uncertainty to signal events in the sideband region, and results in a 3\% uncertainty on the final cross section. 

\begin{figure*}[t]
\centering
\ifnum\DoPrePrint=1
  \subfloat[] {
    \includegraphics[width=0.5\columnwidth]{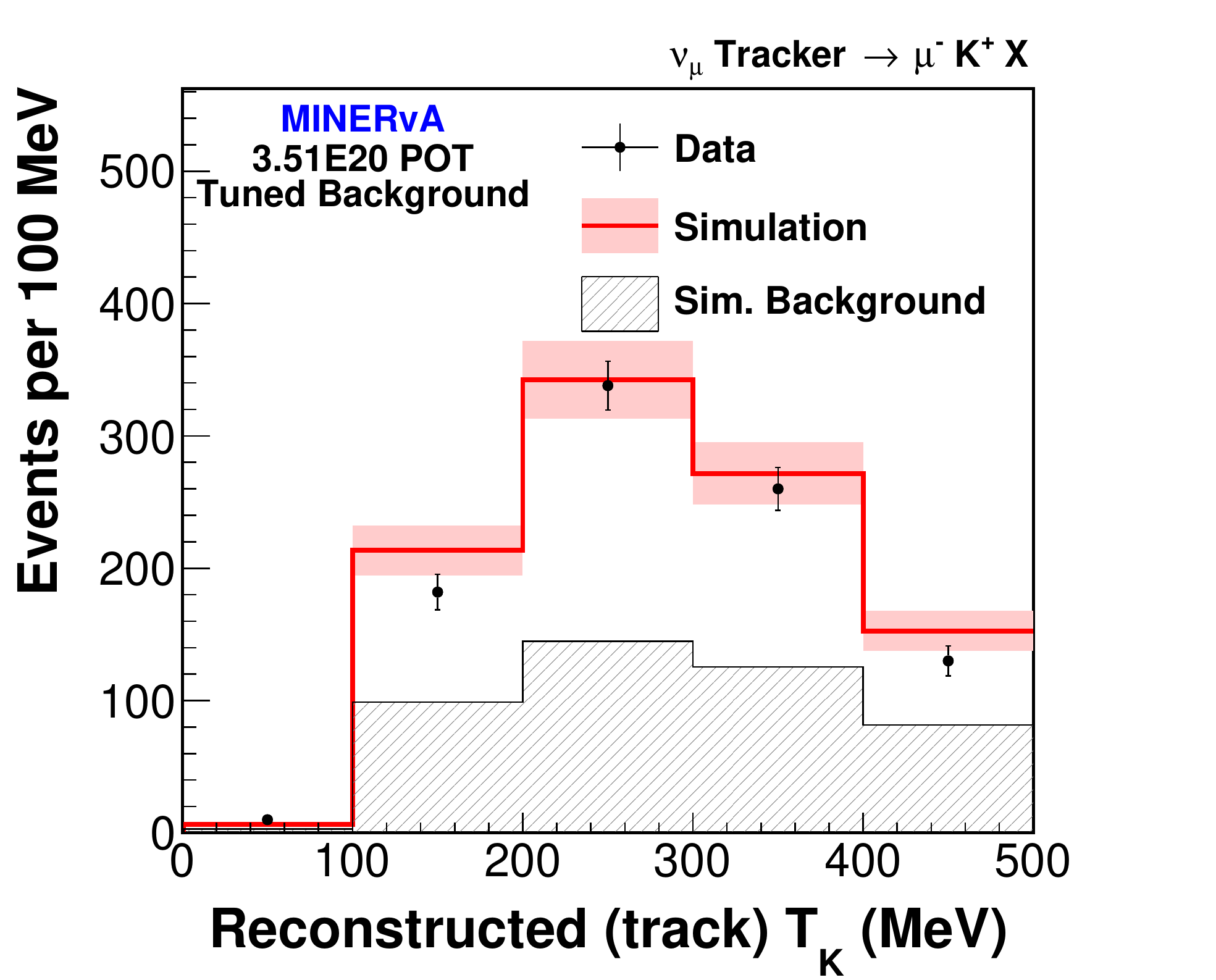}
  }
  \subfloat[] {
    \includegraphics[width=0.5\columnwidth]{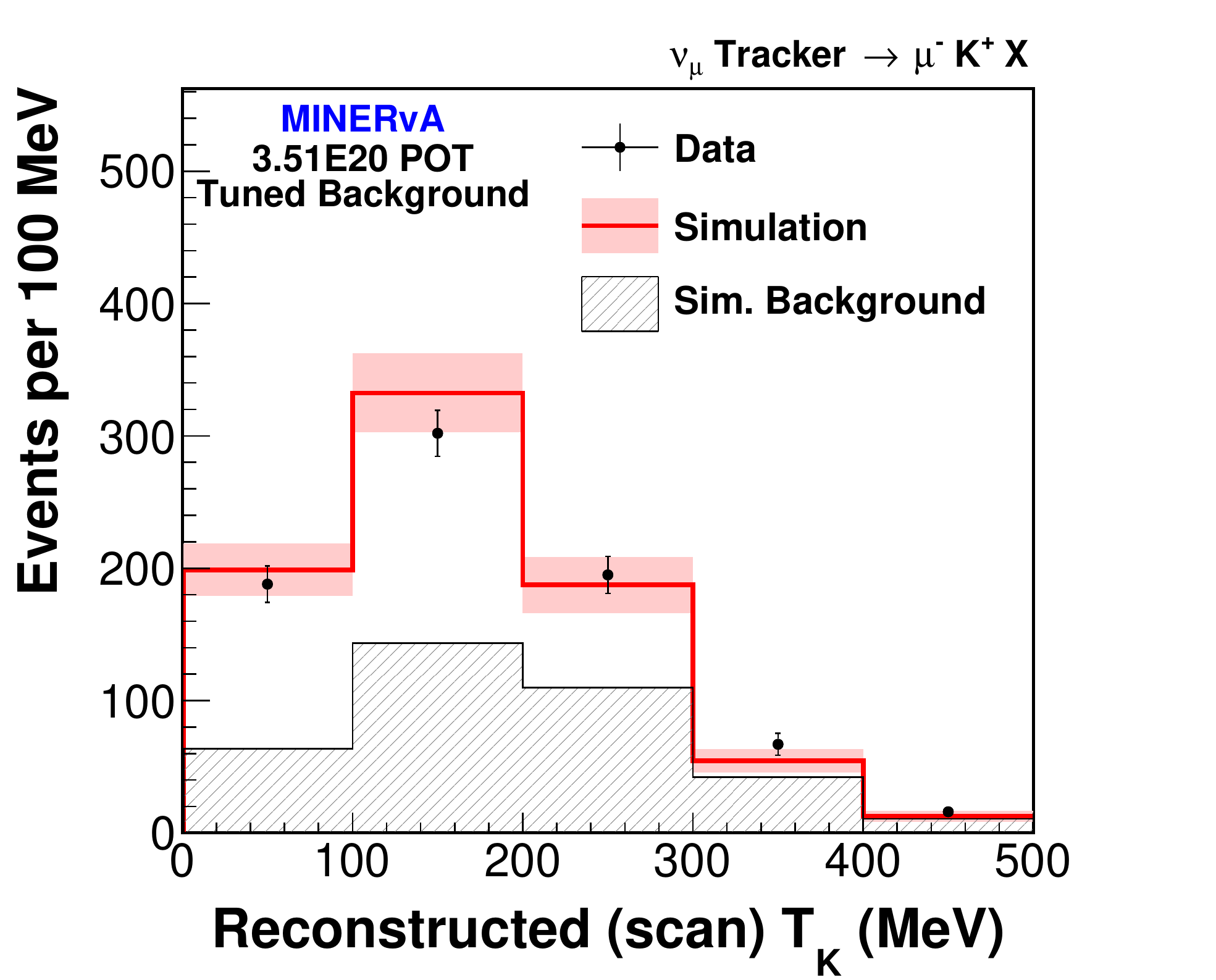}
  }
\else 
  \subfloat[] {
    \includegraphics[width=\columnwidth]{KE_tuned_trk}
  }
  \subfloat[] {
   \includegraphics[width=\columnwidth]{KE_tuned_vtx}
  }
\fi
\caption{(color online) Backgrounds for tracked (left) and untracked (right) events are tuned to data in sidebands and subtracted. \label{fig:bkgsub}}
\end{figure*}

After subtracting backgrounds, and subtracting the estimate of events with true \kp kinetic energy \textgreater~600 MeV, there are 885 signal events in data. Backgrounds are subtracted separately for events reconstructed by tracking and by the event scan, and the background-subtracted samples are then combined. The kinetic energy distributions for selected events with tuned backgrounds are shown for tracked and scanned events in Fig.~\ref{fig:bkgsub}.

\subsection{Unfolding}

\begin{figure}[h]
\centering
\includegraphics[width=\columnwidth]{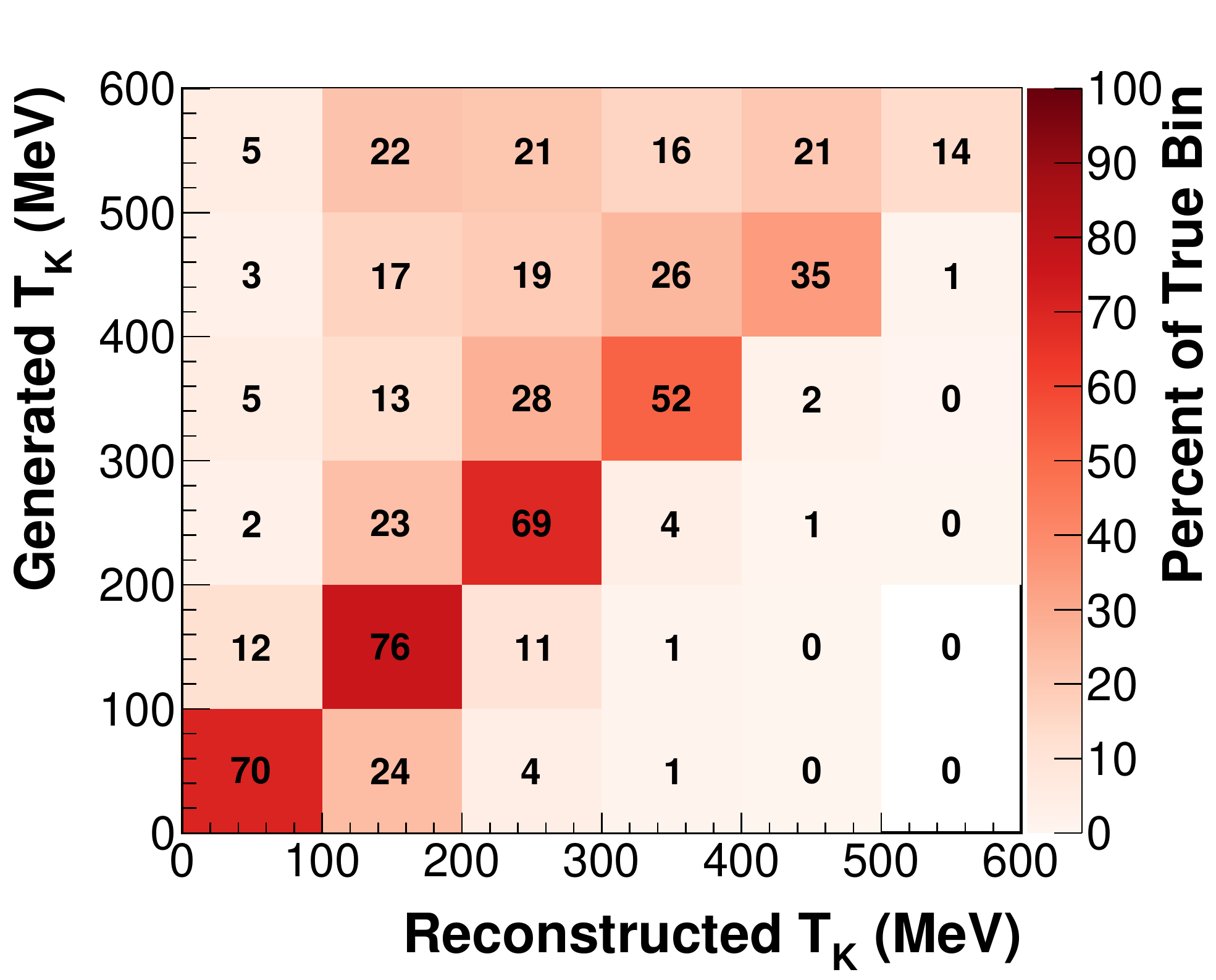}
    \vspace{-7pt}
\caption{(color online) The smearing matrix, normalized so that each each row sums to unity. The number in each cell is the percentage of events of some true kaon kinetic energy that are reconstructed in a given bin. \label{fig:unfolding}}
    \vspace{-10pt}
\end{figure}

The data are unfolded using a Bayesian procedure with three iterations \cite{dagostini}. In addition to correcting for detector resolution effects, the unfolding procedure moves events from low reconstructed kinetic energy to higher true kinetic energy because of hadronic kaon interactions in the detector. These interactions are predicted by Geant4, and reweighted to agree with external measurements of the \kp-carbon elastic and inelastic scattering \cite{bugg,friedman}. The smearing matrix is shown in Fig.~\ref{fig:unfolding}.

The unfolding procedure introduces correlations in the statistical uncertainties. The low and high kinetic energy regions are anticorrelated because of the feed-down from high true kinetic energy to low reconstructed kinetic energy. While we do reconstruct events in the bin from 500 to 600 MeV, 86\% of the true content of that bin smears to lower reconstructed kinetic energy. This bin has a large uncertainty anticorrelated with other bins, and is not reported. A statistical covariance matrix is included in the appendix as Table~\ref{tab:statcov}.

\subsection{Efficiency Correction}

The unfolded distribution is divided by the efficiency, the integrated flux prediction, and the number of nucleons to produce the differential cross section. A correction to the efficiency is calculated due to the lack of strangeness nonconserving reactions such as $\numu n \rightarrow \muminus \kp n$ in the default simulation. The reconstruction efficiency is highest for low-multiplicity final states where the delayed \kp decay products are less likely to deposit energy in scintillator strips which have already been time-stamped with prompt energy due to other particles. Single kaon events have simple final states and thus relatively high efficiency.

An alternative simulated sample is constructed that includes $\Delta S = 1$ reactions. Based on the model of Alam et al.~\cite{atharSK}, these reactions are simulated using \genie version 2.10.0~\cite{genie210} and added to the default simulation, which uses \genie 2.8.4. Events with $\Delta S = 0$ are weighted down by an average of 7.7\% to preserve the total \kp production cross section when the $\Delta S = 1$ events are added. The signal efficiency in this alternate MC is higher in every bin of \kp kinetic energy. The data are corrected by the average of the default and alternate efficiencies, with an uncertainty of 100\% of the correction, such that the error band covers the difference between the efficiency obtained using the two samples. The correction is computed in each bin and is 7.0\% on average.

\subsection{Systematic Uncertainties}

The statistical and systematic uncertainties in each bin are given in Table \ref{tab:syst}. The statistical uncertainty is larger than any single systematic in every bin except for the $400 < T_{K} < 500$~MeV bin. The largest systematics are due to the flux, background model, and \kp interactions in the detector. 

The uncertainty on the integrated flux is 8\%~\cite{MINERvAflux}. The sideband tuning procedure increases the uncertainty on the cross section due to the flux because the high non-kaon hadronic visible energy sideband is from the high-energy tail of the flux, while the signal region at low hadronic energy is mainly from the flux peak. While the dominant effect is from the overall flux normalization, uncertainties in the flux shape enter the analysis through the background subtraction.

The background model uncertainty is dominated by a 100\% uncertainty on pion-carbon interactions inside MINERvA that produce kaons, which are simulated by Geant4 and not constrained by external data. The sideband with high hadronic energy constrains these events, together with high-energy \kp from signal reactions. An uncertainty arises due to the difference in the relative contribution to the signal region and non-kaon hadronic visible energy sideband region from these types of events, which can be seen in Table \ref{tab:bkg}. Uncertainties in the \genie FSI model are evaluated by varying its parameters within measured uncertainties~\cite{genievar1, genievar2}. These variations have little effect on the analysis because the efficiency does not vary strongly with \kp energy, and none of the significant backgrounds depend on the FSI model.

The uncertainty due to kaon interactions includes the effect on the unfolding of varying the \kp-carbon inelastic cross section by $\pm$10\% to cover disagreement between the reweighted Geant4 prediction and external data. It also includes an uncertainty on kaon charge exchange in the detector but outside the struck nucleus. Events where \kz production is followed by $K^{0} p \rightarrow K^{+} n$ are subtracted as background, while there is no acceptance for \kp production events followed by $K^{+} n \rightarrow K^{0} p$. We assign a 100\% uncertainty on both processes and treat it as correlated between the \kp and \kz charge exchange reactions.

The signal model uncertainty is due to the uncertainty in the signal rate for kaons with greater than 600 MeV of kinetic energy, and the uncertainty in the efficiency correction from single kaon production. The rate of high-energy \kp production and the cross section for \kp strong interactions are uncertain. We apply an uncertainty to the high-energy kaons by comparing the ratio of \kp production cross sections above and below 600 MeV using the PYTHIA and KNO hadronization models. The nominal simulation uses KNO for $W$ \textless~ 2.3 GeV, PYTHIA for $W$ \textgreater~ 3.0 GeV, and the AGKY model in between. The resulting additional uncertainty is $_{-11}^{+46}$\% relative to the central value, where KNO, AGKY and PYTHIA are stitched together as a function of $W$.

The energy scale uncertainty comes from two sources. First, an uncertainty of $\pm$6\% is assigned to the energy of the kaon decay product to cover a discrepancy in the peak position in data relative to simulation which can be seen in Fig.~\ref{fig:decayenergy}. Second, uncertainties in the hadronic energy scale affect the non-kaon hadronic visible energy by pushing events from the signal to the sideband region or vice versa. We vary the detector response to hadronic, and electromagnetic, energy based on constraints from a hadron test beam, and a \piz invariant mass peak, respectively. 

The sideband tuning uncertainty is the statistical error on the data in the sideband region, which gives rise to an uncertainty on the scale factor applied in the signal region. The uncertainty due to scanning is dominated by the disagreement between scanners. It also includes a flat 2\% uncertainty because the fraction of events that are rejected in the scan is 2\% higher in simulation than in data. This difference may be due to mismodeling of the relative composition of the sample, which is accounted for by other uncertainties, but is taken as a systematic to be conservative. A summary of statistical and systematic uncertainties is given in Table \ref{tab:syst}.

\begingroup
\squeezetable
\begin{table}
\begin{tabular}{cccccc}
\hline\hline
Source & 0 - 100 & 100 - 200 & 200 - 300 & 300 - 400 & 400 - 500 \\ 
\hline
Statistics & 0.15 & 0.14 & 0.11 & 0.12 & 0.16 \\ 
Flux & 0.09 & 0.11 & 0.10 & 0.11 & 0.13 \\ 
Background model & 0.08 & 0.11 & 0.10 & 0.07 & 0.10 \\ 
Kaon interactions & 0.05 & 0.07 & 0.03 & 0.08 & 0.19 \\ 
Signal model & 0.03 & 0.07 & 0.08 & 0.09 & 0.09 \\ 
Sideband tuning & 0.01 & 0.05 & 0.04 & 0.05 & 0.07 \\ 
Energy scale & 0.02 & 0.02 & 0.03 & 0.04 & 0.05 \\ 
Scanning & 0.05 & 0.05 & 0.04 & 0.03 & 0.03 \\ 
\hline
Total & 0.21 & 0.24 & 0.21 & 0.22 & 0.33 \\ 
\hline\hline
\end{tabular}
\caption{Fractional statistical and systematic uncertainties are reported in bins of kaon kinetic energy, expressed in MeV.}
\label{tab:syst}
\end{table}
\endgroup

\section{Results}
\label{sec:results}

\begin{figure}[h]
\centering
\includegraphics[width=\columnwidth]{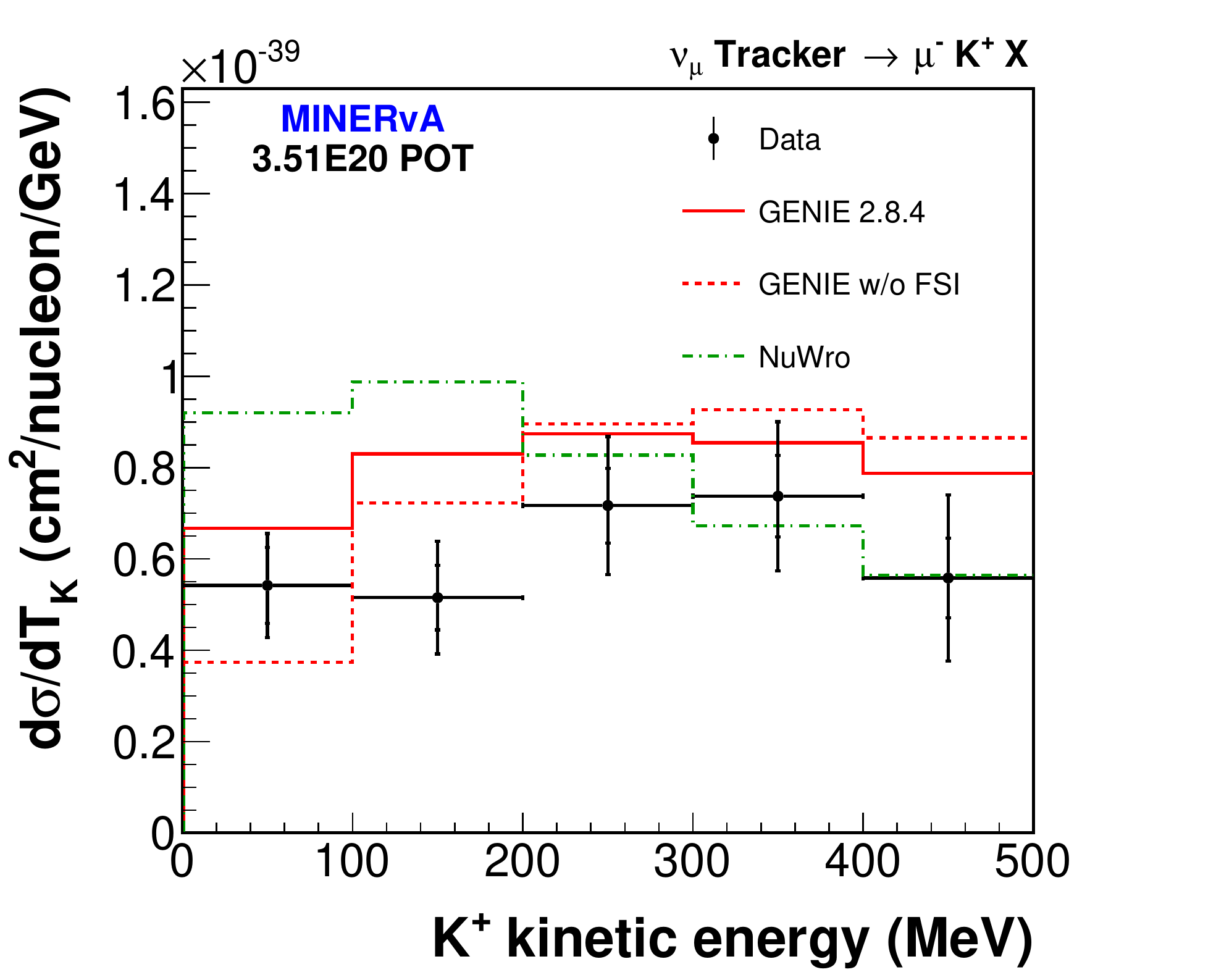}
    \vspace{-7pt}
\caption{(color online) The differential cross section is compared to predictions from the \genie and NuWro event generators. The dashed red line is the \genie prediction with the FSI model turned off. While the simulation used to extract the cross section is reweighted based on \kp interaction fate, the \genie prediction shown here is not reweighted. \label{fig:xs}}
    \vspace{-10pt}
\end{figure}

The extracted differential cross section with respect to the kaon kinetic energy is shown in Fig.~\ref{fig:xs}, along with predictions from \genie 2.8.4 with and without FSI, and the NuWro generator~\cite{nuwro, nuwroHP}. Our data agree best with \genie with FSI. The $\chi^{2}$s for 5 degrees-of-freedom for \genie with FSI, \genie without FSI, and NuWro are 8.1, 11.2, and 27.0, respectively. The shape-only $\chi^{2}$s for 4 degrees-of-freedom are 3.5, 7.8, and 13.1.

In \genie, the nucleon-level cross section is tuned to inclusive \kz and \lam production data on deuterium. Strangeness nonconserving $\Delta S = 1$ events are not simulated in this prediction, but the rate of $\Delta S = 0$ production is tuned to data that does not distinguish between the two. With the addition of a $\Delta S = 1$ component in \genie 2.10~\cite{genie210}, the $\Delta S = 0$ should be reduced, as our data lie 15\% below the prediction. In NuWro, kaon production is not tuned to data. Kaons are produced only in hadronization using PYTHIA for all hadronic invariant masses, and are not subject to FSI. 

The shape of our data agree well with the \genie prediction with final-state interactions. Rescattering, which moves events to lower kinetic energies, is the only channel included in \genie 2.8.4 for $K^{+}$, and improves the agreement with our data significantly compared to the prediction without FSI. The kaon FSI model in \genie lacks both kaon charge exchange and kaon production by pion reactions in the nucleus. Notwithstanding, comparisons with \genie are important due to its widespread use in the neutrino scattering community. \genie is the default simulation for DUNE, and this comparison is the first benchmark of \kp production and \kp FSI.

The addition of charge exchange would decrease the rate of final-state \kp in an isoscalar nucleus like carbon because charged kaons outnumber neutral kaons by 50\% in charged-current interactions in \genie. Kaon production by pion reactions would enhance the cross section. The GiBUU nuclear transport model~\cite{gibuu} predicts a dramatic increase in low-energy kaons due to FSI processes. Either such an enhancement is actually not very large, or the nucleon-level production of \kp would have to be modified downward even further to compensate and still describe the data.

In conclusion, we have made the first high-statistics measurement of the \kp energy spectrum for kaon production in \numu charged-current interactions, with approximately 50 times more events than have been observed in previous experiments~\cite{anl2, anl12ft, bnl7ft, gargamelle}. This result provides a constraint on strange particle production by neutrinos that complements existing measurements of \kzS and \lam production in bubble chambers~\cite{bebc1, bebc2, bebc3, bebc4, fnal15a, fnal15b} and NOMAD~\cite{nomad}. It provides an additional constraint on \kp FSI, suggesting that modifications to the signal spectrum in \pknu due to kaon rescattering are well-modeled in \genie. The disagreement with NuWro illustrates the importance of an improved low-$W$ DIS model for \kp production.

\begin{acknowledgments}

This work was supported by the Fermi National Accelerator Laboratory
under US Department of Energy contract No. DE-AC02-07CH11359 which
included the \minerva\ construction project.  Construction support was
also granted by the United States National Science Foundation under
Award PHY-0619727 and by the University of Rochester. Support for
participating scientists was provided by NSF and DOE (USA), by CAPES
and CNPq (Brazil), by CoNaCyT (Mexico), by CONICYT (Chile), by
CONCYTEC, DGI-PUCP and IDI/IGI-UNI (Peru), and by Latin American
Center for Physics (CLAF).  We thank the
MINOS Collaboration for use of its near detector data. We acknowledge
the dedicated work of the Fermilab staff responsible for the operation
and maintenance of the NuMI beamline, MINERvA and MINOS detectors and
the physical and software environments that support scientific
computing at Fermilab.

\end{acknowledgments}

\newpage
\section{Appendix}

This appendix contains tables of measured cross sections, uncertainties, and bin correlations for the measurement presented in the paper. The correlations are important when comparing this measurement to other predictions, and are taken into account in the computation of the $\chi^{2}$ values for model comparisons given in the results.

\begingroup
\squeezetable
\begin{table}[h]
\begin{tabular}{ccccc}
\hline\hline
$T_{K}$ (MeV) & $d\sigma/dT_{K}$ & Total & Statistical & Systematic \\
\hline
0 - 100 & 0.54 & 0.11 & 0.08 & 0.08 \\ 
100 - 200 & 0.52 & 0.12 & 0.07 & 0.10 \\ 
200 - 300 & 0.72 & 0.15 & 0.08 & 0.13 \\ 
300 - 400 & 0.74 & 0.16 & 0.09 & 0.14 \\ 
400 - 500 & 0.56 & 0.18 & 0.09 & 0.16 \\ 

\hline\hline
\end{tabular}
\caption{The differential cross section with respect to \kp kinetic energy $T_{K}$ is given in units of $10^{-39} cm^{2}$ per nucleon per GeV, as well as the total statistical and systematic uncertainties. A breakdown of the systematic uncertainty is given in Table \ref{tab:bkg}.}
\label{tab:xs}
\end{table}
\endgroup

\begingroup
\squeezetable
\begin{table}[h]
\begin{tabular}{cccccc}
\hline\hline
$T_{K}$ (MeV) ($\times 10^{-80}$) &   0 - 100 & 100 - 200 & 200 - 300 & 300 - 400 & 400 - 500 \\
\hline
  0 - 100 & 0.253 & 0.263 & 0.356 & 0.371 & 0.310 \\
100 - 200 & 0.263 & 0.329 & 0.415 & 0.442 & 0.398 \\
200 - 300 & 0.356 & 0.415 & 0.557 & 0.571 & 0.503 \\
300 - 400 & 0.371 & 0.442 & 0.571 & 0.614 & 0.544 \\
400 - 500 & 0.310 & 0.398 & 0.503 & 0.544 & 0.506 \\
\hline\hline
\end{tabular}
\caption{The covariance for the flux uncertainty.}
\label{tab:nonfluxcov}
\end{table}
\endgroup

\begingroup
\squeezetable
\begin{table}[h]
\begin{tabular}{cccccc}
\hline\hline
$T_{K}$ (MeV) ($\times 10^{-80}$) &   0 - 100 & 100 - 200 & 200 - 300 & 300 - 400 & 400 - 500 \\
\hline
  0 - 100 & 0.347 & 0.367 & 0.405 & 0.192 & 0.040 \\
100 - 200 & 0.367 & 0.694 & 0.673 & 0.425 & 0.248 \\
200 - 300 & 0.405 & 0.673 & 1.050 & 0.814 & 0.728 \\
300 - 400 & 0.192 & 0.425 & 0.814 & 1.248 & 1.395 \\
400 - 500 & 0.040 & 0.248 & 0.728 & 1.395 & 2.042 \\
\hline\hline
\end{tabular}
\caption{The summed covariance for all systematic uncertainties except for the flux. The largest of these are due to background modelling and \kp interactions in the detector.}
\label{tab:statcov}
\end{table}
\endgroup

\begingroup
\squeezetable
\begin{table}
\begin{tabular}{cccccc}
\hline\hline
$T_{K}$ (MeV) ($\times 10^{80}$) &   0 - 100 & 100 - 200 & 200 - 300 & 300 - 400 & 400 - 500 \\
\hline
  0 - 100 & 0.693 & 0.023 & -0.072 & -0.042 & -0.056 \\
100 - 200 & 0.023 & 0.498 & 0.021 & -0.070 & -0.043 \\
200 - 300 & -0.072 & 0.021 & 0.669 & -0.009 & -0.083 \\
300 - 400 & -0.042 & -0.070 & -0.009 & 0.801 & 0.066 \\
400 - 500 & -0.056 & -0.043 & -0.083 & 0.066 & 0.755 \\
\hline\hline
\end{tabular}
\caption{The statistical covariance is nonzero due to the unfolding procedure, which introduces small negative correlations in the statistical uncertainty from bin to bin.}
\label{tab:matrix}
\end{table}
\endgroup

\end{document}